\documentclass[12pt]{article}
\usepackage{eqsection,subeqnarray,indent,amsfonts}

\begin{document}

\title{\vspace*{-2cm}Bounds on the Complex Zeros\\
       of (Di)Chromatic Polynomials\\
       and Potts-Model Partition Functions}
 
\author{
  \\[-1mm]
  {Alan D. Sokal}                   \\
  {\it Department of Physics}       \\
  {\it New York University}         \\
  {\it 4 Washington Place}          \\
  {\it New York, NY 10003 USA}      \\
  {\tt SOKAL@NYU.EDU}          \\
  \\
}
\vspace{0.5cm}
 
\date{April 11, 1999 \\ Version 2: June 18, 1999 \\
      Version 3: October 25, 2000 \\[1mm]
      to appear in {\em Combinatorics, Probability \& Computing}\/}
\maketitle
\thispagestyle{empty}   

\newcommand{\be}{\begin{equation}}
\newcommand{\ee}{\end{equation}}
\newcommand{\<}{\langle}
\renewcommand{\>}{\rangle}
\newcommand{\widebar}{\overline}
\def\reff#1{(\protect\ref{#1})}
\def\spose#1{\hbox to 0pt{#1\hss}}
\def\ltapprox{\mathrel{\spose{\lower 3pt\hbox{$\mathchar"218$}}
 \raise 2.0pt\hbox{$\mathchar"13C$}}}
\def\gtapprox{\mathrel{\spose{\lower 3pt\hbox{$\mathchar"218$}}
 \raise 2.0pt\hbox{$\mathchar"13E$}}}
\def\textprime{${}^\prime$}
\def\proof{\par\medskip\noindent{\sc Proof.\ }}
\def\qed{\hbox{\hskip 6pt\vrule width6pt height7pt depth1pt \hskip1pt}\bigskip}
\def\proofof#1{\bigskip\noindent{\sc Proof of #1.\ }}
\def\half{ {1 \over 2} }
\def\third{ {1 \over 3} }
\def\twothird{ {2 \over 3} }
\def\smfrac#1#2{\textstyle{#1\over #2}}
\def\smhalf{ \smfrac{1}{2} }
\newcommand{\imag}{\mathop{\rm Im}\nolimits}
\newcommand{\real}{\mathop{\rm Re}\nolimits}
\newcommand{\sgn}{\mathop{\rm sgn}\nolimits}
\def\hboxscript#1{ {\hbox{\scriptsize\em #1}} }

\def\scra{{\cal A}}
\def\scrc{{\cal C}}
\def\scrf{{\cal F}}
\def\scrg{{\cal G}}
\def\scrl{{\cal L}}
\def\scro{{\cal O}}
\def\scrp{{\cal P}}
\def\scrr{{\cal R}}
\def\scrs{{\cal S}}
\def\scrt{{\cal T}}
\def\scrv{{\cal V}}
\def\scrz{{\cal Z}}

\def\q{{\sf q}}

\newtheorem{theorem}{Theorem}[section]
\newtheorem{proposition}[theorem]{Proposition}
\newtheorem{lemma}[theorem]{Lemma}
\newtheorem{corollary}[theorem]{Corollary}


\newenvironment{sarray}{
          \textfont0=\scriptfont0
          \scriptfont0=\scriptscriptfont0
          \textfont1=\scriptfont1
          \scriptfont1=\scriptscriptfont1
          \textfont2=\scriptfont2
          \scriptfont2=\scriptscriptfont2
          \textfont3=\scriptfont3
          \scriptfont3=\scriptscriptfont3
        \renewcommand{\arraystretch}{0.7}
        \begin{array}{l}}{\end{array}}
 
\newenvironment{scarray}{
          \textfont0=\scriptfont0
          \scriptfont0=\scriptscriptfont0
          \textfont1=\scriptfont1
          \scriptfont1=\scriptscriptfont1
          \textfont2=\scriptfont2
          \scriptfont2=\scriptscriptfont2
          \textfont3=\scriptfont3
          \scriptfont3=\scriptscriptfont3
        \renewcommand{\arraystretch}{0.7}
        \begin{array}{c}}{\end{array}}






 
\def\Z{{\mathbb Z}}
\def\R{{\mathbb R}}
\def\C{{\mathbb C}}

\begin{abstract}
I show that there exist universal constants $C(r) < \infty$
such that, for all loopless graphs $G$ of maximum degree $\le r$,
the zeros (real or complex) of the chromatic polynomial $P_G(q)$
lie in the disc $|q| < C(r)$.
Furthermore, $C(r) \le 7.963907 r$.
This result is a corollary of a more general result on the zeros
of the Potts-model partition function $Z_G(q, \{v_e\})$
in the complex antiferromagnetic regime $|1 + v_e| \le 1$.
The proof is based on a transformation of the
Whitney--Tutte--Fortuin--Kasteleyn representation
of $Z_G(q, \{v_e\})$ to a polymer gas,
followed by verification of the Dobrushin--Koteck\'y--Preiss condition
for nonvanishing of a polymer-model partition function.
I also show that, for all loopless graphs $G$ of second-largest degree
$\le r$, the zeros of $P_G(q)$ lie in the disc $|q| < C(r) + 1$.
Along the way, I give a simple proof of a generalized (multivariate)
Brown-Colbourn conjecture on the zeros of the reliability polynomial
for the special case of series-parallel graphs.
\end{abstract}

\vspace{0.5cm}
 
\clearpage

\section{Introduction}   \label{sec1}

The polynomials studied in this paper arise independently
in graph theory and in statistical mechanics.
It is appropriate, therefore, to begin by explaining each of these contexts.
Specialists in these fields are warned that they will find at least one
(and perhaps both) of these summaries excruciatingly boring;
they can skip them.

Let $G = (V,E)$ be a finite undirected graph\footnote{
   In this paper a ``graph'' is allowed to have
   loops and/or multiple edges unless explicitly stated otherwise.
}
with vertex set $V$ and edge set $E$.
For each positive integer $q$,
let $P_G(q)$ be the number of ways that the vertices of $G$
can be assigned ``colors'' from the set $\{ 1,2,\ldots,q \}$
in such a way that adjacent vertices always receive different colors.
It is not hard to show (see below) that $P_G(q)$ is the restriction
to $\Z_+$ of a polynomial in $q$.
This (obviously unique) polynomial is called the
{\bf chromatic polynomial}\/ of $G$,
and can be taken as the {\em definition}\/ of $P_G(q)$ for
arbitrary real or complex values of $q$.\footnote{
   Two excellent reviews on chromatic polynomials are
   \cite{Read_68,Read_88}.
   An extensive bibliography on chromatic polynomials is \cite{Chia_97}.
}

The chromatic polynomial was introduced in 1912 by Birkhoff \cite{Birkhoff_12}.
The original hope was that study of the real or complex zeros of $P_G(q)$
might lead to an analytic proof of the Four-Color Conjecture
\cite{Ore_67,Saaty_77},
which states that $P_G(4) > 0$ for all loopless planar graphs $G$.
To date this hope has not been realized, although combinatoric proofs
of the Four-Color Theorem have been found
\cite{Appel_77a,Appel_77b,Appel_89,Robertson_97,Thomas_98}.
Even so, the zeros of $P_G(q)$ are interesting in their own right
and have been extensively studied.
Most of the available theorems concern real zeros
\cite{Birkhoff_46,Tutte_70,Woodall_77,Woodall_92,Jackson_93,Woodall_97,%
Thomassen_97,Edwards_98},
but there has been some study (mostly numerical) of complex zeros as well
\cite{Hall_65,Berman_69,Biggs_72,Beraha_79,Beraha_80,Farrell_80,%
Baxter_86,Baxter_87,Read_91,Wakelin_92,Brenti_94,Brown_98a,Brown_98b,%
Shrock_97a,Shrock_97b,Shrock_97c,Shrock_97d,Shrock_98a,Shrock_98b,%
Shrock_98c,Tsai_98,Shrock_98e,Shrock_99a,Shrock_99b,Salas-Sokal_in_prep}.

A more general polynomial can be obtained as follows:
Assign to each edge $e \in E$ a real or complex weight $v_e$.
Then define
\be
   Z_G(q, \{v_e\})   \;=\;
   \sum_{ \{\sigma_x\} }  \,  \prod_{e \in E}  \,
      \biggl[ 1 + v_e \delta(\sigma_{x_1(e)}, \sigma_{x_2(e)}) \biggr]
   \;,
 \label{eq1.1}
\ee
where the sum runs over all maps $\sigma\colon\, V \to \{ 1,2,\ldots,q \}$,
the $\delta$ is the Kronecker delta,
and $x_1(e), x_2(e) \in V$ are the two endpoints of the edge $e$
(in arbitrary order).
It is not hard to show (see below)
that $Z_G(q, \{v_e\})$ is the restriction to $q \in \Z_+$ of
a polynomial in $q$ and $\{v_e\}$.
If we take $v_e = -1$ for all $e$, this reduces to the chromatic polynomial.
If we take $v_e = v$ for all $e$, this defines a two-variable polynomial
$Z_G(q,v)$ that was introduced
implicitly by Whitney \cite{Whitney_32a,Whitney_32b,Whitney_33}
and explicitly by Tutte \cite{Tutte_47,Tutte_54};
it is known variously (modulo trivial changes of variable)
as the {\bf dichromatic polynomial}\/, the {\bf dichromate}\/,
the {\bf Whitney rank function}\/ or the {\bf Tutte polynomial}\/
\cite{Welsh_93,Biggs_93}.\footnote{
   The Tutte polynomial $T_G(x,y)$ is conventionally defined as
   \protect\cite[p.~45]{Welsh_93} \protect\cite[pp.~73, 101]{Biggs_93}
   $$
     T_G(x,y)   \;=\;
     \sum\limits_{E' \subseteq E} (x-1)^{k(E')-k(E)} \,
                                (y-1)^{|E'| + k(E') - |V|}
   $$
   where $k(E')$ is the number of connected components
   in the subgraph $(V,E')$.
   Comparison with (\protect\ref{eq1.2}) below yields
   $$
     T_G(x,y)   \;=\;   (x-1)^{-k(E)} \, (y-1)^{-|V|} \,
                        Z_G \bigl( (x-1)(y-1), \, y-1 \bigr)   \;.
   $$
}

In statistical mechanics, \reff{eq1.1} is known as the
partition function of the
{\bf \mbox{\boldmath $q$}-state Potts model}\/.\footnote{
   The Potts model \cite{Potts_52} was invented in the early 1950's by Domb
   (see \cite{Domb_74}).
   The $q=2$ case, known as the Ising model \cite{Ising_25},
   was invented in 1920 by Lenz \cite{Lenz_20}
   (see \cite{Brush_67,Kobe_97,Ising_obit}).
   The $q=4$ case, which is a special case of the Ashkin--Teller model,
   was invented in 1943 by Ashkin and Teller \cite{Ashkin-Teller_43}.
}
In the Potts model \cite{Wu_82,Wu_84},
an ``atom'' (or ``spin'') at site $x \in V$ can exist in any one of
$q$ different states (where $q$ is an integer $\ge 1$).
The {\bf energy}\/ of a configuration is the sum, over all edges $e \in E$,
of $0$ if the spins at the two endpoints of that edge are unequal
and $-J_e$ if they are equal.
The {\bf Boltzmann weight}\/ of a configuration is then $e^{-\beta H}$,
where $H$ is the energy of the configuration
and $\beta \ge 0$ is the inverse temperature.
The {\bf partition function}\/ is the sum, over all configurations,
of their Boltzmann weights.
Clearly this is just a rephrasing of \reff{eq1.1},
with $v_e = e^{\beta J_e} - 1$.
A coupling $J_e$ (or $v_e$)
is called {\bf ferromagnetic}\/ if $J_e \ge 0$ ($v_e \ge 0$)
and {\bf antiferromagnetic}\/ if $-\infty \le J_e \le 0$ ($-1 \le v_e \le 0$).

To see that $Z_G(q, \{v_e\})$ is indeed a polynomial in its arguments
(with coefficients that are in fact 0 or 1), we proceed as follows:
In \reff{eq1.1}, expand out the product over $e \in E$,
and let $E' \subseteq E$ be the set of edges for which the term
$v_e \delta_{\sigma_{x_1(e)}, \sigma_{x_2(e)}}$ is taken.
Now perform the sum over configurations $\{ \sigma_x \}$:
in each connected component of the subgraph $(V,E')$
the spin value $\sigma_x$ must be constant,
and there are no other constraints.
Therefore,
\be
   Z_G(q, \{v_e\})   \;=\;
   \sum_{ E' \subseteq E }  q^{k(E')}  \prod_{e \in E'}  v_e
   \;,
 \label{eq1.2}
\ee
where $k(E')$ is the number of connected components
(including isolated vertices) in the subgraph $(V,E')$.
The expansion \reff{eq1.2} was discovered by
Birkhoff \cite{Birkhoff_12} and Whitney \cite{Whitney_32a}
for the special case $v_e = -1$ (see also Tutte \cite{Tutte_47,Tutte_54});
in its general form it is due to
Fortuin and Kasteleyn \cite{Kasteleyn_69,Fortuin_72}
(see also \cite{Edwards-Sokal}).
We take \reff{eq1.2} as the {\em definition}\/ of $Z_G(q, \{v_e\})$
for arbitrary complex $q$ and $\{v_e\}$.

In statistical mechanics, a very important role is played by the
complex zeros of the partition function.
This arises as follows \cite{Yang-Lee_52}:
Statistical physicists are interested in {\em phase transitions}\/,
namely in points where one or more physical quantities
(e.g.\ the energy or the magnetization)
depend nonanalytically (in many cases even discontinuously)
on one or more control parameters
(e.g.\ the temperature or the magnetic field).
Now, such nonanalyticity is manifestly impossible in \reff{eq1.1}/\reff{eq1.2}
for any finite graph $G$.
Rather, phase transitions arise only in the {\em infinite-volume limit}\/.
That is, we consider some countably infinite graph
$G_\infty = (V_\infty, E_\infty)$
--- usually a regular lattice, such as $\Z^d$ with nearest-neighbor edges ---
and an increasing sequence of finite subgraphs $G_n = (V_n, E_n)$.
It can then be shown
(under modest hypotheses on the $G_n$)
that the {\bf (limiting) free energy per unit volume}\/
\be
   f_{G_\infty}(q,v)   \;=\;
   \lim_{n \to \infty}   |V_n|^{-1}  \log Z_{G_n}(q,v)
 \label{limiting_free_energy}
\ee
exists for all {\em nondegenerate physical}\/ values
of the parameters\footnote{
   Here ``physical'' means that the weights are nonnegative,
   so that the model has a probabilistic interpretation;
   and ``nondegenerate'' means that we exclude the limiting cases
   $v=-1$ in (a) and $q=0$ in (b), which cause difficulties
   due to the existence of configurations having zero weight.
},
namely either
\begin{quote}
\begin{itemize}
  \item[(a)]     $q$ integer $\ge 1$ and $-1 < v < \infty$
     \quad  [using \reff{eq1.1}:  see e.g.\ \cite[Section I.2]{Israel_79}]
  \item[or (b)]  $q$ real $> 0$ and $0 \le v < \infty$
     \quad  [using \reff{eq1.2}:  see \cite[Theorem 4.1]{Grimmett_95}
                                  and \cite{Grimmett_78,Seppalainen_98}].
\end{itemize}
\end{quote}
This limit $f_{G_\infty}(q,v)$ is in general a continuous function of $v$;
but it can fail to be a real-analytic function of $v$,
because complex singularities of $\log Z_{G_n}(q,v)$
--- namely, complex zeros of $Z_{G_n}(q,v)$ ---
can approach the real axis in the limit $n \to\infty$.
Therefore, the possible points of physical phase transitions
are precisely the real limit points of such complex zeros.
As a result, theorems that constrain the possible location of
complex zeros of the partition function are of great interest.
In particular, theorems guaranteeing that a certain complex domain
is free of zeros are often known as {\em Lee-Yang theorems}\/.\footnote{
   The first such theorem, concerning the behavior of the ferromagnetic
   Ising model at complex magnetic field, was proven by Lee and Yang
   \cite{Lee-Yang_52} in 1952.
   A partial bibliography (through 1980) of generalizations of this result
   can be found in \cite{Lieb-Sokal_81}.
}

The purpose of this paper is to prove an upper bound on the
complex $q$-plane zeros of the
Potts-model partition function $Z_G(q, \{v_e\})$,
valid throughout the ``complex antiferromagnetic regime'' $|1 + v_e| \le 1$,
under certain ``local'' conditions on the weights $\{v_e\}$:
for example, in terms of the quantity
$\max\limits_{x \in V} \sum\limits_{e \ni x} |v_e|$.
As a corollary, I obtain upper bounds
on the zeros of the chromatic polynomial $P_G(q)$
in terms of the maximum degree of the graph $G$.
More precisely, I show that there exist universal constants $C(r) < \infty$
such that, for all loopless graphs $G$ of maximum degree $\le r$,
the zeros of $P_G(q)$ lie in the disc $|q| < C(r)$.
This answers in the affirmative a question posed by
Brenti, Royle and Wagner \cite[Question 6.1]{Brenti_94},
generalizing an earlier conjecture of
Biggs, Damerell and Sands \cite{Biggs_72}
limited to $r$-regular graphs.
The constants $C(r)$ arise as the solution of an explicit minimization problem,
and I prove that $C(r) \le 7.963907 r$.
This linear dependence on $r$ is best possible,
as the example of the complete graph $K_{r+1}$ shows that $C(r) \ge r$.

Furthermore, I show that the presence of {\em one}\/ vertex
of large degree cannot lead to large chromatic roots.
More precisely, if all but one of the vertices of $G$ have degree $\le r$,
then the zeros of $P_G(q)$ lie in the disc $|q| < C(r) + 1$.
Please note that a result of this kind {\em cannot}\/ hold if ``all but one''
is replaced by ``all but two'', for in this case the chromatic roots
can be unbounded, even when $r=2$ and $G$ is planar \cite{Sokal_hierarchical}.

The proofs of these results are based on well-known methods
of mathematical statistical mechanics.
The first step is to transform the
Whitney--Tutte--Fortuin--Kasteleyn representation \reff{eq1.2}
into a gas of ``polymers'' interacting via a hard-core exclusion
(Section \ref{sec2}).
I then invoke the Dobrushin condition \cite{Dobrushin_96a,Dobrushin_96b}
(or the closely related Koteck\'y--Preiss condition 
 \cite{Kotecky_86,Sokal_Mayer_in_prep})
for the nonvanishing of a polymer-model partition function
(Section \ref{sec3}).
Lastly, I verify these conditions for our particular polymer model,
using a series of simple combinatorial lemmas,
some of which may be of independent interest (Section \ref{sec4});
in particular, I give a simple proof of a generalized (multivariate)
Brown-Colbourn conjecture on the zeros of the reliability polynomial
for the special case of series-parallel graphs
(Remark 3 in Section \ref{sec4.1}).
The main results of this paper are contained in Section \ref{sec5};
some generalizations and extensions are in Section \ref{sec6}.
I conclude with some conjectures and open questions (Section \ref{sec7}).

With a little more work,
it should be possible to extend the arguments of this paper
to prove the existence and analyticity
of the limiting free energy per unit volume \reff{limiting_free_energy}
for suitable regular lattices $G_\infty$
and translation-invariant edge weights $v_e$,
in the same region of complex $q$- and $\{v_e\}$-space
where $Z$ will be proven (in Section~\ref{sec5})
to be nonvanishing uniformly in the finite subgraphs $V_n$
(``uniformly in the volume'' in statistical-mechanical lingo).
In particular, this would provide a convergent expansion for
the limiting free energy in powers of $1/q$.
However, I have not worked out the details.

This paper would never have seen the light of day
without the help and advice of Antti Kupiainen.
During my visit to Helsinki in September--October 1997,
I told Antti of my conjectures about $P_G(q)$ and $Z_G(q, \{v_e\})$
--- conjectures that I had no good idea how to prove.
He immediately saw that they ought to be provable by
cluster (or Mayer) expansion.
My reaction was, ``Ugh!  You know how I {\em detest}\/ the cluster
expansion!'';
indeed, I had resisted learning it for nearly 20 years
and had devoted much of my work in mathematical physics
to finding ways of circumventing it
\cite{Sokal_82,BFS_83a,BFS_83b,FFS_book}.
Antti assured me that the cluster expansion is not so difficult,
and he suggested that I study the
excellent review article of Brydges \cite{Brydges_86}.
We also quickly figured out how to represent
$Z_G(q, \{v_e\})$ as a polymer gas.
Jean Bricmont then told me about the work of Koteck\'y and Preiss
\cite{Kotecky_86},
and Roman Koteck\'y informed me of the work of Dobrushin \cite{Dobrushin_96a}.
Here, finally, was a version of the cluster expansion simple enough
that even I could understand it!
Nine months later, I figured out how to verify the
Dobrushin (or Koteck\'y--Preiss) condition and thereby complete the proof.

\section{Transformation of the Potts-Model Partition Function
         to a Polymer Gas}   \label{sec2}

Let $G=(V,E)$ be a finite undirected graph
equipped with complex edge weights $\{ v_e \}_{e \in E}$.
If $G$ contains a loop $e$ (i.e.\ an edge connecting a vertex to itself),
this simply multiplies $Z_G(q, \{v_e\})$ by a factor $1+v_e$;
so we can assume without loss of generality that $G$ is loopless,
and we shall do so in this section in order to avoid unnecessary complications.
Likewise, if $G$ contains multiple edges $e_1,\ldots,e_n$
connecting the same pair of vertices, they can be replaced,
without changing the value of $Z$,
by a single edge $e$ with weight $v_e = \prod_{i=1}^n (1+v_{e_i}) - 1$.
So we could assume without loss of generality, if we wanted,
that $G$ has no multiple edges.
But this assumption would not simplify most of our subsequent arguments,
so we shall usually refrain from making it.
Note, however, that our numerical bounds frequently get better
if multiple edges are replaced by a single equivalent edge.

So let $G$ be loopless, and consider the
Whitney--Tutte--Fortuin--Kasteleyn representation \reff{eq1.2}
of the Potts-model partition function $Z_G(q, \{v_e\})$.
For each term in \reff{eq1.2}
we decompose the subgraph $(V,E')$ into its connected components.
Some of these components may consist of a single vertex and no edges;
the remaining components are disjoint connected subgraphs
$(S_1,E_1), \ldots, (S_N,E_N)$ with $|S_i| \ge 2$.
The total number of components is
\begin{subeqnarray}
   k(E')   & = &  N \,+\, \left( |V| - \sum_{i=1}^N |S_i| \right)  \\[2mm]
           & = &  |V|  \,-\, \sum_{i=1}^N (|S_i| - 1)
\end{subeqnarray}
It follows that:

\begin{proposition}[jointly with Antti Kupiainen]
   \label{prop2.1}
Let $G=(V,E)$ be a loopless finite undirected graph
equipped with edge weights $\{ v_e \}_{e \in E}$.
Then
\be
   Z_G(q, \{v_e\})   \;=\;   q^{|V|} Z_{polymer,G}(q, \{v_e\})   \;,
  \label{eq2.2}
\ee
where
\be
   Z_{polymer,G}(q, \{v_e\})   \;=\;
   \sum_{N=0}^\infty  {1 \over N!} \;
   \sum_{S_1,\ldots,S_N \,\hboxscript{disjoint}} \;
   \prod_{i=1}^N w(S_i)
  \label{eq2.3}
\ee
and
\be
   w(S)   \;=\;
   \cases{ q^{-(|S|-1)}  \!\!
           \sum\limits_{\begin{scarray}
                            \widetilde{E} \subseteq E  \\
                            (S,\widetilde{E}) \hboxscript{ connected}
                        \end{scarray}}
           \prod\limits_{e \in \widetilde{E}} v_e
           & if $|S| \ge 2$   \cr
           \noalign{\vskip 2mm}
           0   &  if $|S| \le 1$
         }
  \label{eq2.4}
\ee
The sum in \reff{eq2.3} runs over pairwise disjoint subsets
$S_1,\ldots,S_N$ of $V$,
and the term $N=0$ in \reff{eq2.3} is understood to contribute 1.
\end{proposition}

\smallskip
\par\noindent
Note, in particular, that $w(S) = 0$ if $S$ is disconnected
[i.e.\ if the induced subgraph $(S, E_S)$ is disconnected].

The ``polymer model'' \reff{eq2.3}--\reff{eq2.4}
has the form of a grand-canonical gas
(see Section~\ref{sec3} for the precise definition)
\be
   Z_{polymer,G}(q, \{v_e\})   \;=\;
   \sum_{N=0}^\infty  {1 \over N!}
   \sum_{S_1,\ldots,S_N}  \prod_{i=1}^N w(S_i)
                          \prod_{1 \le i < j \le N}  W(S_i,S_j)
  \label{eq2.5}
\ee
with single-particle state space $\scrp_*(V)$
[the set of all nonempty subsets of $V$],
fugacities $w(S)$,
and two-particle Boltzmann factor given by a hard-core exclusion
\be
   W(S,S')   \;=\;
   \cases{ 1  & if $S \cap S' = \emptyset$  \cr
           \noalign{\vskip 1mm}
           0  & otherwise \cr
         }
  \label{eq2.6}
\ee
Graph theorists will recognize the right-hand side of \reff{eq2.5}
as the generating function, in the variables $w(S)$,
for independent subsets of vertices of the
intersection graph of $\scrp_*(V)$.

The usefulness of \reff{eq2.2}--\reff{eq2.6}
comes from the fact that the fugacities $w(S)$
are all suppressed by powers of $q^{-1}$,
hence are small for large $|q|$.
Moreover, if the sum over $\widetilde{E}$ in \reff{eq2.4} can be controlled,
one expects that $w(S)$ will be exponentially decaying in $|S|$
when $|q|$ is large enough.
This raises the hope that the Mayer expansion \cite{Uhlenbeck_62},
which is an expansion of $\log Z_{polymer,G}$
in powers of the fugacities $w(S)$,
might converge for sufficiently large $|q|$.
If so, this would imply that $Z_{polymer,G} \neq 0$
in the region of convergence.
That is what we go about proving in the following sections ---
but in the opposite order.

\section{Dobrushin and Koteck\'y--Preiss Conditions for\break\hfill
         the Nonvanishing of $Z$}    \label{sec3}

In statistical mechanics, a {\bf grand-canonical gas}
is defined by a {\em single-particle state space}\/ $X$
(here assumed for simplicity to be finite),
a {\em fugacity vector}\/ $w = \{w_x\}_{x \in X} \in \C^X$,
and a {\em two-particle Boltzmann factor}\/ $W(x,y)$
[a symmetric function $W \colon\; X \times X \to \C$].
The (grand) partition function $Z(w,W)$ is then defined to be the sum
over ways of placing $N \ge 0$ ``particles''
on ``sites'' $x_1, \ldots, x_N \in X$,
with each configuration assigned a ``Boltzmann weight''
given by the product of the corresponding factors
$w_{x_i}$ and $W(x_i,x_j)$:
\be
   Z(w,W)  \;=\;  \sum_{N=0}^\infty {1 \over N!}  \;
      \sum_{x_1,\ldots,x_N \in X}  \;
           \prod_{i=1}^N w_{x_i}
           \prod_{1 \le i < j \le N}  W(x_i,x_j)
   \;,
 \label{eq3.1}
\ee
where the $N=0$ term is understood to contribute 1.
Under very mild conditions on $W$
[e.g.\ $|W(x,y)| \le 1$ for all $x,y$ is more than sufficient],
$Z(w,W)$ is an entire analytic function of $w$.
Our goal is to find a sufficient condition for $Z(w,W)$
to be nonvanishing in a polydisc $D_R = \{ w \colon\;  |w_x| < R_x\}$.
This would imply, in particular, that $\log Z(w,W)$
is an analytic function of $w$ in $D_R$.

We say that $W$ is
\begin{itemize}
   \item  {\em physical}\/ if $0 \le W(x,y) < +\infty$ for all $x,y \in X$
   \item  {\em repulsive}\/ if $|W(x,y)| \le 1$ for all $x,y \in X$
   \item  {\em physical and repulsive}\/ if $0 \le W(x,y) \le 1$
     for all $x,y \in X$
   \item  {\em hard-core}\/ if $W(x,y) = 0 \hbox{ or } 1$
     for all $x,y \in X$
   \item  {\em hard-core self-repulsive}\/ if $W(x,x) = 0$
     for all $x \in X$
\end{itemize}
An important special case is when $W$ is
hard-core and hard-core self-repulsive:
then $Z(w,W)$ is the generating function for independent sets of vertices
of the graph $\widetilde{G} = (X,E)$ defined by placing an edge
between each pair of vertices $x \neq y$ for which $W(x,y) = 0$.

Dobrushin \cite{Dobrushin_96a,Dobrushin_96b}
has given an elegant sufficient condition
for the nonvanishing of $Z$ in a polydisc $D_R$,
whenever $W$ is hard-core and hard-core self-repulsive.
His proof is astoundingly simple,
avoiding all the combinatoric complication that has
given cluster expansions such a reputation for difficulty.
Here I shall present a slight extension of Dobrushin's theorem,
in which the condition of hard-core interaction is replaced
by the weaker assumption that the interaction is physical and repulsive;
moreover, the conclusion of the theorem is slightly strengthened.
(We won't really need this extension ---
the original Dobrushin theorem would suffice for our purposes ---
but the stronger result is no more difficult,
and it gives a bit more insight into the method of proof.)
The hard-core {\em self-}\/repulsion is, however, essential
both in Dobrushin's version and in my own:
it guarantees that each ``site'' $x \in X$ can be occupied
by at most one ``particle'' $x_i$.
It follows that the partition function can be rewritten
as a sum over subsets:
\be
   Z(w,W)  \;=\;  \sum_{X' \subseteq X}  \;
           \prod_{x \in X'} w_x
           \prod_{\< xy \> \in X'}   W(x,y)
   \;,
\ee
where the second product runs over unordered pairs $x,y \in X'$
($x \neq y$) with each pair counted once.

Let us define, for each subset $\Lambda \subseteq X$,
the restricted partition function
\be
   Z_\Lambda(w,W)  \;=\;  \sum_{X' \subseteq \Lambda}  \;
           \prod_{x \in X'} w_x
           \prod_{\< xy \> \in X'}   W(x,y)
   \;.
   \label{def_Z_Lambda}
\ee
Of course this notation is redundant,
since the same effect can be obtained by setting $w_x = 0$
for $x \in X \setminus \Lambda$,
but it is useful for the purposes of the inductive proof.
We have:

\begin{theorem}
  \label{thm3.1}
Let $X$ be a finite set, and let $W$ satisfy
\begin{itemize}
   \item[(a)]  $0 \le W(x,y) \le 1$ for all $x,y \in X$
   \item[(b)]  $W(x,x) = 0$ for all $x \in X$
\end{itemize}
Suppose there exist constants $R_x \ge 0$ and $0 \le K_x < 1/R_x$
satisfying
\be
   K_x  \;\ge\;  \prod_{y \neq x}   {1 - W(x,y) K_y R_y \over  1 - K_y R_y}
  \label{eq1}
\ee
for all $x \in X$.
Then, for each subset $\Lambda \subseteq X$,
$Z_\Lambda(w,W)$ is nonvanishing in the closed polydisc
$\bar{D}_R = \{ w \in \C^X \colon\; |w_x| \le R_x \}$
and satisfies there
\be
  \left|  {\partial\log Z_\Lambda (w,W) \over \partial w_x} \right|
  \;\le\;
  \cases{ {\displaystyle {K_x \over 1 - K_x |w_x|} }
                              & for all $x \in \Lambda$  \cr
           \noalign{\vskip 4mm}
          0                   & for all $x \in X \setminus \Lambda$ \cr
        }
 \label{eq2}
\ee
Moreover, if $w, w' \in \bar{D}_R$
and $w'_x / w_x \in [0,+\infty]$ for each $x \in \Lambda$,
then
\be
  \left|  \log {Z_\Lambda(w',W) \over Z_\Lambda(w,W)}  \right|
  \;\le\;  \sum_{x \in \Lambda}
  \left|  \log {1 - K_x |w'_x|  \over  1 - K_x |w_x|}  \right|
 \label{eq3}
\ee
where on the left-hand side we take the standard branch of the log,
i.e.\ $|\imag\log \cdots| \le \pi$.
\end{theorem}

\medskip
\par\noindent
{\bf Remarks.}  1. It follows from \reff{eq1} that $K_x \ge 1$
and hence that $R_x < 1$.

2. The conclusion of Dobrushin's theorem \cite{Dobrushin_96a,Dobrushin_96b}
is the special case of \reff{eq3}
in which some of the $w'_x$ are equal to $w_x$
and others are equal to 0,
and in which only the {\em real part}\/ of the logarithm
on the left-hand side is handled.

\medskip

\proof
Note first that \reff{eq2} for any given $\Lambda$ implies \reff{eq3}
for the same $\Lambda$, by integration.

The proof is by induction on the cardinality of $\Lambda$.
If $\Lambda = \emptyset$ the claims are trivial.
So let us assume that \reff{eq2} [and hence also \reff{eq3}] holds for
all sets of cardinality $< n$,
and let a set $\Lambda$ of cardinality $n$ be given.
Let $x$ be any element of $\Lambda$,
and let $\Lambda' = \Lambda \setminus \{x\}$.
It follows from \reff{def_Z_Lambda} that
\be
   Z_\Lambda(w,W) \;=\;
   Z_{\Lambda'}(w,W)  \,+\, w_x Z_{\Lambda'}(\widetilde{w},W)
  \label{eq_basic}
\ee
where
\be
   \widetilde{w}_y  \;=\;  W(x,y) \, w_y  \;;
\ee
here the first term on the right-hand side of \reff{eq_basic} covers the
summands $X' \not\ni x$, while the second covers $X' \ni x$.
Note that $\widetilde{w} \in \bar{D}_R$ since $|W(x,y)| \le 1$.
{}From \reff{eq_basic} we have
\be
   {\partial \over \partial w_x} \log Z_\Lambda(w,W)
   \;=\;
   {k(w)  \over  1 + k(w) w_x}
\ee
where
\be
   k(w)  \;=\;  {Z_{\Lambda'}(\widetilde{w},W)  \over  Z_{\Lambda'}(w,W)}  \;.
\ee
Now by the inductive hypothesis \reff{eq3} for $\Lambda'$,
and using the fact that $\widetilde{w}_y/w_y = W(x,y) \ge 0$, we have
\be
   |k(w)|  \;\le\;
   \prod_{y \in \Lambda'}
       {1 - W(x,y) K_y |w_y| \over  1 - K_y |w_y|}
   \;\le\;
   \prod_{y \in X \setminus \{x\}}
       {1 - W(x,y) K_y |w_y| \over  1 - K_y |w_y|}
   \;,
\ee
which is $\le K_x$ by the hypothesis \reff{eq1}.
This proves \reff{eq2} for $\Lambda$, and hence completes the induction.
\qed

Let us now return to the special case of a hard-core interaction.
If $W(x,y) = 0$ (resp.\ 1), we say that $x$ and $y$ are
{\em incompatible}\/ (resp.\ {\em compatible}\/)
and write $x \not\sim y$ (resp.\ $x \sim y$).
Note that in our convention $x \not\sim x$,
in agreement with some authors' convention
\cite{Kotecky_86,Seiler_82,Simon_93}
and contrary to others' \cite{Dobrushin_96a,Dobrushin_96b}.
The hypothesis \reff{eq1} is then equivalent to the existence
of constants $c_x \ge 0$ such that
\be
   R_x  \;\le\;
   (e^{c_x} - 1) \, \exp\!\left( - \sum_{y \not\sim x} c_y \right)
 \label{dob_condition}
\ee
for all $x \in X$  [set $c_x = -\log(1 - K_x R_x)$].
This is the Dobrushin \cite{Dobrushin_96a,Dobrushin_96b} condition.
Slightly stronger, and more convenient to check, is the
Koteck\'y--Preiss \cite{Kotecky_86,Sokal_Mayer_in_prep} condition
\be
   R_x  \;\le\;
   c_x \, \exp\!\left( - \sum_{y \not\sim x} c_y \right)
   \;.
 \label{KP_condition}
\ee

\bigskip

Let us now consider the important special case in which
the single-particle state space $X$ can be partitioned as
$X = \bigcup\limits_{n=1}^\infty X_n$ in such a way that
\be
   \sum\limits_{y \in X_n \colon\; y \not\sim x}   R_y
   \;\le\;  A_n m
   \quad\hbox{for all $x \in X_m$}
 \label{partition_condition_1}
\ee
for suitable constants $\{A_n\}_{n=1}^\infty$.
[This typically arises when
$X$ is some set of nonempty subsets of a finite set $V$,
and $x \not\sim y$ means $x \cap y \neq \emptyset$;
we will then take $X_n$ to be the sets of cardinality $n$,
and will prove \reff{partition_condition_1}
by proving
\be
   \sum\limits_{y \in X_n \colon\; y \ni i}   R_y
   \;\le\;  A_n
   \quad\hbox{for all $i \in V$}
   \;,
 \label{partition_condition_2}
\ee
which is manifestly stronger than \reff{partition_condition_1}.]
Let us take
\be
    c_x \;=\; e^{\alpha n} R_x   \quad\hbox{for all }  x \in X_n
 \label{partition_choice_cx}
\ee
with some suitably chosen $\alpha > 0$.
Then, for \reff{partition_condition_1} to imply the
Koteck\'y--Preiss condition \reff{KP_condition},
it suffices that
\be
   \sum\limits_{n=1}^\infty e^{\alpha n} \, A_n   \;\le\; \alpha   \;.
 \label{KP_sum_condition}
\ee
We have therefore proven:

\begin{proposition}
   \label{prop3.2}
Suppose that $X = \bigcup\limits_{n=1}^\infty X_n$ (disjoint union)
and that there exist constants $\{A_n\}_{n=1}^\infty$ and $\alpha$
such that
\begin{itemize}
   \item[(a)]  $\sum\limits_{y \in X_n \colon\; y \not\sim x}   R_y
      \;\le\;  A_n m
      \quad\hbox{for all $m,n$ and all $x \in X_m$}$
   \item[(b)]  $\sum\limits_{n=1}^\infty e^{\alpha n} \, A_n   \;\le\; \alpha
                \;.$
\end{itemize}
Then the Koteck\'y--Preiss condition \reff{KP_condition}
holds with the choice $c_x = e^{\alpha n} R_x$ for $x \in X_n$.
\end{proposition}

\medskip
\par\noindent
{\bf Remarks.}  1.  Suppose we try the more general Ansatz
$c_x = b_n R_x$ for $x \in X_n$.  Then \reff{partition_condition_1}
implies the Koteck\'y--Preiss condition \reff{KP_condition} in case
$b_n \ge e^{\alpha n}$ where $\alpha \equiv \sum_{n=1}^\infty b_n A_n$.
But in that case $b'_n \equiv e^{\alpha n} \ge e^{\alpha' n}$
where $\alpha' \equiv \sum_{n=1}^\infty b'_n A_n$.
So there is no loss of generality in restricting attention
to $b_n = e^{\alpha n}$ for some $\alpha$.

2.  Since the state space $X$ is finite, only finitely many of the $A_n$
are nonzero.  Nevertheless, we often have occasion to consider
simultaneously an infinite {\em family}\/ of problems --- e.g.\ in this paper,
all loopless graphs $G$ of maximum degree $\le r$ and arbitrarily many vertices
--- and it is natural to seek bounds that are {\em uniform}\/
over the family.  So it is useful to forget that only finitely many
of the $A_n$ are nonzero.  (Moreover, similar methods can be applied
to problems with an infinite state space $X$, in which case $\{A_n\}$
is a genuinely infinite sequence.)  This leads to two further remarks:

3.  For the condition
\be
   \exists \alpha > 0 \hbox{ such that }
   \sum\limits_{n=1}^\infty e^{\alpha n} A_n   \;\le\; \alpha
  \label{KP_sum_condition_2}
\ee
to hold, it is necessary that the sequence $\{ A_n \}_{n=1}^\infty$
have {\em some}\/ exponential decay
(i.e.\ $A_n \le C e^{-\epsilon n}$ for some $\epsilon > 0$),
but there is no minimum required rate of decay.
Indeed, if $\{ A_n \}_{n=1}^\infty$ has any exponential decay at all,
then by modifying finitely many of the $A_n$ one can make
\reff{KP_sum_condition_2} hold.
It can thus be valuable in applications to work hard on estimating
the first few coefficients $A_n$
(see \cite{Kennedy_88} for an example).\footnote{
   The emphasis in
   \protect\cite{Cammarota_82,Seiler_82,Brydges_86,Kotecky_86,Simon_93}
   on the special case $A_n = C e^{-\epsilon n}$ with $C=1$
   is thus somewhat misleading, inasmuch as it suggests
   that there is a minimum allowed rate of decay $\epsilon$.
}

4.  Let $\delta = \liminf_{n\to\infty} (-\log A_n)/n$.
Then $F(\alpha) = \alpha^{-1} \sum_{n=1}^\infty e^{\alpha n} A_n$
is finite-valued and continuous (in fact, real-analytic)
on $0 < \alpha < \delta$,
left-continuous (as a map into the extended real line)
as $\alpha \uparrow \delta$,
and identically $+\infty$ for $\alpha > \delta$.
In particular, the infimum of $F(\alpha)$ is attained,
so \reff{KP_sum_condition_2} is equivalent to
\be
   \inf\limits_{\alpha > 0}
      \alpha^{-1} \sum\limits_{n=1}^\infty e^{\alpha n} A_n
   \;\le\;   1 \;.
\ee

\bigskip
\par\noindent
{\bf Important Final Remark.}
The results in this section provide an extraordinarily simple proof
of the convergence of the Mayer expansion
for a grand-canonical gas with physical and repulsive
two-particle interactions.
To see what is at issue, let us first trivially rewrite
the partition function \reff{eq3.1} as
\be
   Z(w,W)  \;=\;  \sum_{N=0}^\infty {1 \over N!}  \;
      \sum_{x_1,\ldots,x_N \in X}  \;
           \prod_{i=1}^N w_{x_i}
           \sum_{G \in \scrg_N}  \prod_{\<ij\> \in G}   F(x_i,x_j)
   \;,
\ee
where $\scrg_N$ is the set of all (simple loopless undirected) graphs
on the vertex set $\{1,\ldots,N\}$,
and
\be
   F(x,y)  \;=\;  W(x,y) - 1
\ee
is called the {\em two-particle Mayer factor}\/.
Then standard combinatorial arguments \cite{Uhlenbeck_62}
show that
\be
   \log Z(w,W)  \;=\;  \sum_{N=1}^\infty {1 \over N!}  \;
      \sum_{x_1,\ldots,x_N \in X}  \;
           \prod_{i=1}^N w_{x_i}
           \sum_{G \in \scrc_N}  \prod_{\<ij\> \in G}   F(x_i,x_j)
 \label{mayer_exp}
\ee
at least in the sense of formal power series in $w$,
where $\scrc_N \subseteq \scrg_N$ is the set of {\em connected}\/ graphs
on $\{1,\ldots,N\}$.
This is the Mayer expansion;
the principal problem is to prove its convergence in
some specified polydisc.
The usual approach to proving convergence of the Mayer expansion
\cite{Penrose_67,Seiler_82,Cammarota_82,Brydges_86,Brydges_87,Brydges_88,%
Simon_93,Brydges_99,Sokal_Mayer_in_prep}
is to explicitly bound the terms in \reff{mayer_exp};
this requires some rather nontrivial combinatorics
(for example, Proposition~\ref{prop_penrose} below
 together with the counting of trees).
Once this is done, an immediate consequence is that $Z$ is nonvanishing
in any polydisc where the series for $\log Z$ is convergent.
Dobrushin's brilliant idea \cite{Dobrushin_96a,Dobrushin_96b}
was to prove these two results in the opposite order.
First one proves, by an elementary induction on the cardinality
of the state space, that $Z$ is nonvanishing in some specified polydisc
(Theorem~\ref{thm3.1});
it then follows immediately that $\log Z$ is analytic in that polydisc,
and hence that its Taylor series \reff{mayer_exp} is convergent there.
It is an interesting open question to know whether this approach
can be made to work without the assumption of hard-core self-repulsion.

\section{Some Combinatorial Lemmas}   \label{sec4}

\subsection{Reduction to trees}  \label{sec4.1}

The weight $w(S)$ involves a sum \reff{eq2.4}
over connected subgraphs $(S,\widetilde{E})$
of the induced subgraph $(S, E_S)$.
The trouble is that there may be ``too many'' connected subgraphs.
It is remarkable, therefore, that this sum can sometimes be bounded
by a sum over a much smaller set of graphs, namely {\em spanning trees}\/.
The following proposition underlines the special role played by the
``complex antiferromagnetic regime''
$A \equiv \{v \in \C \colon\; |1 + v| \le 1 \}$.

\begin{proposition}[Penrose \protect\cite{Penrose_67}]
  \label{prop_penrose}
Let $G=(V,E)$ be a finite undirected graph
equipped with complex edge weights $\{ v_e \}_{e \in E}$
satisfying $|1 + v_e| \le 1$ for all $e$.
Then
\be
   \left|  \sum\limits_{\begin{scarray}
                            E' \subseteq E  \\
                            (V,E') \hboxscript{ connected}
                        \end{scarray}}
           \, \prod\limits_{e \in E'} v_e \,
   \right|\,
   \;\,\le\;\,
           \sum\limits_{\begin{scarray}
                            E' \subseteq E  \\
                            (V,E') \hboxscript{ tree}
                        \end{scarray}}
           \, \prod\limits_{e \in E'} |v_e|
   \;.
 \label{penrose_bound}
\ee
\end{proposition}

\par\noindent
Penrose \cite{Penrose_67} proved this when $G$ is the complete graph $K_n$;
the result then follows for all graphs without loops or multiple edges
(it suffices to set $v_e = 0$ on the nonexistent edges).
Here I present a minor modification of Penrose's proof
that permits loops and multiple edges:

\medskip

\proof
We can assume without loss of generality that $G$ is connected,
since otherwise both sides of the inequality are zero.
Let $\scrc$ (resp.\ $\scrt$) be the set of subsets $E' \subseteq E$
such that $(V,E')$ is connected (resp.\ is a tree).
Clearly $\scrc$ is an increasing family of subsets of $E$
with respect to set-theoretic inclusion,
and the minimal elements of $\scrc$ are precisely those of $\scrt$
(i.e.\ the spanning trees).
It is a nontrivial but well-known fact
\cite[Sections 7.2 and 7.3]{Bjorner_92}
\cite[Section 8.3]{Ziegler_95}
that the (anti-)complex $\scrc$ is {\em partitionable}\/:
that is, there exists a map ${\bf R} \colon\, \scrt \to \scrc$
such that ${\bf R}(T) \supseteq T$ for all $T \in \scrt$
and $\scrc = \biguplus \, [T, \, {\bf R}(T)]$ (disjoint union),
where $[E_1,E_2]$ denotes the Boolean interval
$\{ E' \colon\; E_1 \subseteq E' \subseteq E_2 \}$.
In fact, many alternative choices of ${\bf R}$ are available
\cite[Sections 7.2 and 7.3]{Bjorner_92} 
\cite[Sections 2 and 6]{Gessel_96}
\cite[Proposition 13.7 et seq.]{Biggs_93},
and none of the subsequent arguments will depend on a specific choice
of ${\bf R}$.  Nevertheless, for completeness, we shall give at the end
of this proof a concrete construction of one possible ${\bf R}$.

Given the existence of ${\bf R}$, we have the immediate identity
\begin{eqnarray}
   \sum\limits_{\begin{scarray}
                   E' \subseteq E  \\
                   (V,E') \hboxscript{ connected}
                \end{scarray}}
   \; \prod\limits_{e \in E'} v_e
   & = &
   \sum\limits_{\begin{scarray}
                   T \subseteq E  \\
                   (V,T) \hboxscript{ tree}
                \end{scarray}}
   \; \prod\limits_{e \in T} v_e
   \sum\limits_{T \subseteq E' \subseteq {\bf R}(T)}
   \; \prod\limits_{e \in E' \setminus T} v_e
      \nonumber \\[3mm]
   & = &
   \sum\limits_{\begin{scarray}
                   T \subseteq E  \\
                   (V,T) \hboxscript{ tree}
                \end{scarray}}
   \; \prod\limits_{e \in T} v_e
   \; \prod\limits_{e \in {\bf R}(T) \setminus T} (1 + v_e)
   \;.
 \label{penrose_identity}
\end{eqnarray}
In particular, if $|1 + v_e| \le 1$ for all $e$,
then \reff{penrose_bound} follows.

We now indicate a construction of ${\bf R}$
that is a slight variant of the one used by Penrose \cite{Penrose_67}
(he orders the vertices, while I order the edges):
Choose (arbitrarily) a vertex $x \in V$ and call it the root;
and choose (arbitrarily) a numbering of the edges.
For each $E' \in \scrc$ and $y \in V$, let $\hbox{\em depth}_{E'}(y)$
be the length of the shortest path in $E'$ connecting $y$ to the root.
For each $y \in V \setminus \{x\}$, let $e(y)$ be the lowest-numbered edge
in $E'$ connecting $y$ to a vertex $y'$ with
$\hbox{\em depth}_{E'}(y') = \hbox{\em depth}_{E'}(y) - 1$.
And finally, let
${\bf S}(E') = \{ e(y) \colon\; y \in V \setminus \{x\} \, \}$.
Then trivially ${\bf S}(E') \subseteq E'$;
moreover, it is easy to see that $(V, {\bf S}(E'))$ is a tree
and that $\hbox{\em depth}_{{\bf S}(E')}(y) = \hbox{\em depth}_{E'}(y)$
for all $y \in V$.
Conversely, given a spanning tree $(V,T)$, it is not hard to see that
${\bf S}(E') = T$ if and only if $T \subseteq E' \subseteq {\bf R}(T)$,
where ${\bf R}(T)$ is obtained from $T$ by adjoining all edges $e \in E$ that
\begin{itemize}
   \item[(a)]  connect two vertices of equal $\hbox{\em depth}_{T}$
      (this includes loops, if any), or
   \item[(b)]  connect a vertex $y$ to a vertex $y'$ having
      $\hbox{\em depth}_{T}(y') = \hbox{\em depth}_{T}(y) - 1$
      where $e$ is higher-numbered than the edge already in $T$
      that connects $y$ to a vertex $y''$ having
      $\hbox{\em depth}_{T}(y'') = \hbox{\em depth}_{T}(y) - 1$.
\end{itemize}
This completes the proof.
\qed

{\bf Remarks.}
1.  The identity \reff{penrose_identity} and
the inequality \reff{penrose_bound} generalize to matroids.
Indeed, for any matroid $M$, the independent sets of $M$ form
a simplicial complex $IN(M)$, called a {\em matroid complex}\/;
moreover, every matroid complex is shellable,
and every shellable complex is partitionable
\cite[Theorem 7.3.3 and Proposition 7.2.2]{Bjorner_92}.
For the cographic matroid $M^*(G)$,
the independent sets are the complements of connected subgraphs,
and the bases are complements of spanning trees,
so we recover the situation of Proposition~\ref{prop_penrose}.
I thank Criel Merino for teaching me about matroids
and for helping me notice a silly error
in my original proof of Proposition~\ref{prop_penrose}.
Earlier, Dave Wagner had informed me that analogues of \reff{penrose_identity}
hold for shellable simplicial complexes
(see \cite[sections 0.3 and III.2]{Stanley_96} for the definition).
There is a long history of identities related to \reff{penrose_identity}:
see, for example,
\cite{Rota_64,Malyshev_79,Brydges_86,Brydges_87,Brydges_88,Abdesselam_95,%
Brydges_99}
and the references cited therein.

2.  I conjecture that \reff{penrose_bound} can be strengthened so that
on the right-hand side the absolute value is put outside the sum
rather than inside.
(This would be useful in case the $\{v_e\}$ do not all have the same phase.)
In fact, I conjecture more:
Let
\be
   c(E')   \;=\;   |E'| \,-\, |V| \,+\, k(E')
\ee
be the cyclomatic number of the subgraph $(V,E')$,
and define the generalized connected sum
\begin{subeqnarray}
   C_G(\lambda, \{v_e\})   & = &
   \!\!\sum\limits_{\begin{scarray}
                       E' \subseteq E  \\
                       (V,E') \hboxscript{ connected}
                    \end{scarray}}
   \!\lambda^{c(E')} \, \prod\limits_{e \in E'} v_e
      \\[2mm]
   & = &  \lambda^{1-|V|} \, C_G(1, \{\lambda v_e\})
      \slabel{eq4.4b}  \\[2mm]
   & = & \lim\limits_{q\to 0} \, q^{-1} \lambda^{-|V|}
         Z_G(\lambda q, \{\lambda v_e\})
         \;.
\end{subeqnarray}
In particular, $\lambda=0$ corresponds to the tree sum
and $\lambda=1$ to the connected sum.
Then I conjecture that
\begin{quote}
\begin{itemize}
   \item[(a)]  If $|1 + v_e| \le 1$ for all $e$,
       then $|C_G(\lambda, \{v_e\})|$ is a decreasing function of $\lambda$
       on $0 \le \lambda \le 1$.
\end{itemize}
\end{quote}
I had originally conjectured a stronger result, namely
\begin{quote}
\begin{itemize}
   \item[($\hbox{a}'_r$)]  If $|r + v_e| \le r$ for all $e$,
       then $(-1)^n \, (d^n/d\lambda^n) \, |C_G(\lambda, \{v_e\})|^2 \ge 0$
       on $0 \le \lambda \le 1$, for all $n \ge 0$
\end{itemize}
\end{quote}
either for $r=1$ or, failing that, for $r = {1 \over 2}$;
but this is in fact false for all $r > 0$,
even for the second derivative evaluated at $\lambda = 0$
with equal edge weights $v_e = v$.
Indeed, if we write
\be
   C_G(\lambda,v)   \;=\;   v^{|V|-1} \,
         \bigl[ a_0 + a_1 v \lambda + a_2 v^2 \lambda^2 + \ldots \bigr]
\ee
where $a_j$ is the number of spanning subgraphs of $G$ having $j$ cycles,
then \break
$(d^2 / d\lambda^2) \, |C_G(\lambda,v)|^2 \, \Bigr| _{\lambda=0} \ge 0$
holds for all $v$ if $a_1^2 \ge 2 a_0 a_2$,
but fails for $v$ in a wedge near the imaginary axis if $a_1^2 < 2 a_0 a_2$.
Now the complete bipartite graph $K_{3,4}$
has $a_0 = 432$, $a_1 = 612$, $a_2 = 456$
and hence provides a counterexample.
Nevertheless, ($\hbox{a}'_r$) might be true for some interesting
subclasses of graphs $G$.

For {\em real}\/ $v_e \in [-1,0]$, by contrast, I am able to prove a result
even stronger than ($\hbox{a}'_{1/2}$), namely
\begin{quote}
\begin{itemize}
   \item[(b)]  If $-1 \le v_e \le 0$ for all $e$,
       then $(-1)^{n+|V|-1} \, (d^n/d\lambda^n) \, C_G(\lambda, \{v_e\}) \ge 0$
       on $0 \le \lambda \le 1$, for all integers $n \ge 0$.
\end{itemize}
\end{quote}
Indeed, by \reff{penrose_identity} and \reff{eq4.4b}, we have
\be
   C_G(\lambda, \{v_e\})   \;=\;
   \sum\limits_{\begin{scarray}
                   T \subseteq E  \\
                   (V,T) \hboxscript{ tree}
                \end{scarray}}
   \; \prod\limits_{e \in T} v_e
   \; \prod\limits_{e \in {\bf R}(T) \setminus T} (1 + \lambda v_e)
\ee
and hence
\be
   {d^n \over d\lambda^n} \, C_G(\lambda, \{v_e\})   \;=\;
   \sum\limits_{\begin{scarray}
                   T \subseteq E  \\
                   (V,T) \hboxscript{ tree}
                \end{scarray}}
   \sum\limits_{\begin{scarray}
                   \widetilde{T} \subseteq
                        \hbox{\bf\scriptsize R}(T) \setminus T \\
                   |\widetilde{T}| = n
                \end{scarray}}
   \; \prod\limits_{e \in T \cup \widetilde{T}} v_e
   \; \prod\limits_{e \in {\bf R}(T) \setminus (T \cup \widetilde{T})}
                                                        (1 + \lambda v_e)
   \;,
\ee
which has the claimed sign whenever $0 \le \lambda \le 1$
and $-1 \le v_e \le 0$ for all $e$.

3.  $C_G(1, \{v_e\})$ is equal, up to a prefactor,
to the reliability polynomial $R_G(\{p_e\})$ \cite{Colbourn_87},
where $p_e$ is the probability that edge $e$ is operational
and $v_e = p_e/(1-p_e)$:
\be
   R_G(\{p_e\})  \;=\;   \left[ \prod_{e \in E} (1-p_e) \right]
                         C_G(1, \{p_e/(1-p_e)\})
   \;.
\ee
%
%
%
Now the Brown-Colbourn conjecture \cite{Brown_92,Wagner_00}
states that for any connected graph $G$
(loops and multiple edges are allowed),
$R_G(p) \neq 0$ whenever $|p-1| > 1$.
A more general conjecture is that $R_G(\{p_e\}) \neq 0$
whenever $|p_e - 1| > 1$ for all edges $e$,
or equivalently, that $C_G(1, \{v_e\}) \neq 0$
whenever $0 < |1 + v_e| < 1$ for all $e$.
But this generalized Brown-Colbourn conjecture
is an immediate consequence of conjecture (a):
for if we had $C_G(1, \{v_e\}) = 0$ with $|1 + v_e| < 1$ for all $e$,
then we could choose $\epsilon > 0$ such that $v'_e \equiv (1+\epsilon) v_e$
satisfy $|1 + v'_e| < 1$ for all $e$, and we would have
$C_G(\lambda, \{v'_e\}) = 0$ for $\lambda = 1/(1+\epsilon)$
[but not, of course, identically for $1/(1+\epsilon) \le \lambda \le 1$].

Note also that if the generalized Brown-Colbourn conjecture
holds for a graph $G$, then it holds also for any graph that can be
obtained from $G$ by a sequence of
doublings of edges (``parallel expansions'')
and/or subdivisions of edges (``series expansions'').
This follows from the formulae \cite[p.~35]{Colbourn_87}
\begin{eqnarray}
   R_{G'}(\{p_e, p_1, p_2\})   & = &
       R_G(\{p_e, 1 - (1-p_1)(1-p_2)\})      \\[2mm]
   R_{G'}(\{p_e, p_1, p_2\})   & = &  [1- (1-p_1)(1-p_2)] \,
       R_G \Bigl( \Bigl\{ p_e, {\displaystyle p_1 p_2
                                \over
                                \displaystyle p_1 + p_2 - p_1 p_2
                               } \Bigr\} \Bigr)
\end{eqnarray}
where $G'$ is obtained from $G$ by parallel (resp.\ series) expansion
of an edge $e_0$ into a pair of edges $e_1, e_2$.
It suffices to note that if $|1-p_i| > 1$ for $i=1,2$,
then the same inequality holds for
$p_{\parallel} \equiv 1- (1-p_1)(1-p_2)$
and for $p_{series} \equiv p_1 p_2/(p_1 + p_2 - p_1 p_2)$;
the former is obvious, and the latter follows by observing that
the series-expansion formula corresponds to addition of $1/v = 1/p - 1$
and that $|1-p| > 1$ corresponds to $\real(1/v) < -1/2$.
In particular, since the generalized Brown-Colbourn conjecture manifestly holds
for trees, it also holds for all connected graphs without a $K_4$ minor,
as these are precisely the graphs that can be obtained from trees
by a sequence of series and parallel expansions
\cite{Duffin_65,Liu_76,Wald_83,Oxley_86}.
The (original) Brown-Colbourn conjecture for series-parallel graphs
was first proven by Wagner \cite{Wagner_00},
by a vastly more complicated method.

\subsection{Connected subgraphs containing a specified vertex}

Let $G=(V,E)$ be a finite or countably infinite undirected graph
equipped with edge weights $\{ v_e \}_{e \in E}$,
and let $x \in V$.
Let us define the weighted sum over connected subgraphs
$G' = (V',E') \subseteq G$
containing $n$ vertices, one of which is $x$, and $m$ edges:
\be
   C_{n,m}(G, \{ v_e \}, x)   \;=\;
   \sum_{\begin{scarray}
            G' = (V',E') \subseteq G \\
            G' \hboxscript{ connected} \\
            V' \ni x \\
            |V'| = n \\
            |E'| = m
         \end{scarray}}
   \prod_{e \in E'} |v_e|
   \;.
\ee
Special cases are the tree sum
\be
   T_n(G, \{ v_e \}, x)   \;=\;  C_{n,n-1}(G, \{ v_e \}, x)
\ee
and the edge-counted sum
\be
   C_{\bullet,m}(G, \{ v_e \}, x)  \;=\;
   \sum\limits_{n=1}^{m+1} C_{n,m}(G, \{ v_e \}, x)
   \;.
\ee
When the edge weights $v_e$ are all equal to 1,
we shall optionally omit them from the notation;
note in particular the obvious bound
\be
   C_{n,m}(G, \{ v_e \}, x)   \;\le\;
   C_{n,m}(G, x)  \, \left( \sup_{e \in E} |v_e| \right) ^{\! m}
   \;.
\ee
In this subsection we shall obtain a variety of upper bounds on
$C_{n,m}(G, \{ v_e \}, x)$
in terms of ``local'' information about the graph $G$
and the weights $\{ v_e \}$.

\begin{proposition}
   \label{prop4.2}
Let $G=(V,E)$ be a finite or countably infinite loopless undirected graph
of maximum degree $\le r$,
equipped with edge weights $\{ v_e \}_{e \in E}$;
and let $x \in V$.
Let ${\bf T}_r$ be the infinite $r$-regular tree,
and let $y$ be any vertex in ${\bf T}_r$.
Then
\be
   C_{\bullet,m}(G, x)  \;\le\;
   C_{\bullet,m}({\bf T}_r, y)  \;=\;
   T_{m+1}({\bf T}_r, y)
   \;\equiv\; t_{m+1}^{(r)}
   \;=\;  r \, {[(r-1)(m+1)]!   \over  m! \, [(r-2)m + r]!}
 \label{eq_tnr}
\ee
and hence
\be
   C_{\bullet,m}(G, \{ v_e \}, x)  \;\le\;
   t_{m+1}^{(r)} \,
   \left( \sup_{e \in E} |v_e| \right) ^{\! m}
   \;.
 \label{eq4.16}
\ee
In particular,
\be
   T_n(G, \{ v_e \}, x)   \;\le\;
   t_n^{(r)} \,
   \left( \sup_{e \in E} |v_e| \right) ^{\! n-1}
   \;.
 \label{eq4.17}
\ee
\end{proposition}

\proof
We can assume without loss of generality that $G=(V,E)$ is connected.
Let $U = (\widetilde{V}, \widetilde{E})$ be the universal covering graph of $G$,
with covering map $f \colon\, U \to G$;
and let $\widetilde{x}$ be a vertex of $U$ such that $f(\widetilde{x}) = x$.
[The universal covering graph of a connected loopless graph $G$
 can be constructed as follows:
 Fix a base vertex $x$ of $G$, and let the vertices of $U$ be the
 walks in $G$ (of finite length) that begin at $x$
 and do not contain any ``doublebacks''
 (i.e.\ two consecutive uses of the same edge in opposite directions).
 Two vertices of $U$ are defined to be adjacent if one of them
 is a one-step extension of the other,
 and $f \colon\, U \to G$ maps each walk onto its final vertex.
 We take $\widetilde{x}$ to be the zero-step walk starting at $x$.]
It is easy to see that $U$ is a tree
(in general countably infinite even when $G$ is finite).
Moreover, since $G$ has maximum degree $\le r$,
$U$ is a subtree of ${\bf T}_r$, from which it follows trivially that
$C_{\bullet,m}(U, \widetilde{x})  \le C_{\bullet,m}({\bf T}_r, \widetilde{x})$.
Let us prove, then, that
$C_{\bullet,m}(G, x)  \le C_{\bullet,m}(U, \widetilde{x})$.

Fix an arbitrary total order on $E$,
and choose arbitrarily for each edge $e \in E$ a distinguished direction.
Now let $H$ be a connected $m$-edge subgraph of $G$ that contains $x$.
Let $S$ be the lexicographically first (with respect to the chosen
total order on $E$) spanning tree of $H$.
Then $S$ based at $x$ has a unique lifting to a subgraph $\widetilde{S}$ of $U$
based at $\widetilde{x}$:  it is defined by mapping each vertex $s$ of $S$
to the unique path in $S$ from $x$ to $s$.

Now, for each edge $e$ of $H$ not belonging to $S$,
there is a unique edge $\widetilde{e}$ of $U$ such that $f(\widetilde{e}) = e$
and $\widetilde{e}$ is incident with the image in $\widetilde{S}$
of the vertex of $S$ from which $e$ is directed.
The addition of these edges to $\widetilde{S}$
produces a connected $m$-edge subgraph $\widetilde{H}$ of $U$
that contains $\widetilde{x}$.
Moreover, the map $H \mapsto \widetilde{H}$ is injective,
since $H = f(\widetilde{H})$.
This completes the proof that
$C_{\bullet,m}(G, x)  \le C_{\bullet,m}({\bf T}_r, \widetilde{x})$.

We conclude by calculating the numbers
$t_{m+1}^{(r)} \equiv C_{\bullet,m}({\bf T}_r, \widetilde{x})$.\footnote{
   For similar computations, see e.g.\ \protect\cite{Fisher_61}.
}
Let ${\bf U}_r$ be the infinite tree in which all vertices have degree $r$
except for one vertex $y$ which has degree $r-1$,
and let $u_{m+1}^{(r)} = C_{\bullet,m}({\bf U}_r, y)$.
Then define, as formal power series, the generating functions
\begin{eqnarray}
   T_r(z)  & = &   \sum\limits_{n=1}^\infty  t_n^{(r)} z^n
      \label{Tr_def}    \\[1mm]
   U_r(z)  & = &   \sum\limits_{n=1}^\infty  u_n^{(r)} z^n
      \label{Ur_def}
\end{eqnarray}
The recursive structure of $r$-regular rooted trees
easily implies the functional equations
\begin{eqnarray}
   T_r(z)  & = &   z [1 + U_r(z)]^r
      \label{Tr_eqn}    \\[1mm]
   U_r(z)  & = &   z [1 + U_r(z)]^{r-1}
      \label{Ur_eqn}
\end{eqnarray}
We now use the Lagrange Implicit Function Theorem for formal power series
\cite[Theorem 1.2.4]{Goulden_83},
which states that for formal power series
$f(u) = \sum_{n=0}^\infty f_n u^n$ and
$g(u) = \sum_{n=0}^\infty g_n u^n$ with $g_0 \neq 0$,
the functional equation $U(z) = z g(U(z))$ has a unique solution $U(z)$,
and for all $n \ge 1$ one has
\be
   [z^n] f(U(z))   \;=\;  {1 \over n} \, [u^{n-1}] (f'(u) g(u)^n)
\ee
where $[z^n] P(z)$ denotes the coefficient of $z^n$ in the formal power
series $P(z)$.
Applying this with $f(u) = (1+u)^r$ and $g(u) = (1+u)^{r-1}$ yields
\begin{eqnarray}
   t_n^{(r)}   & = &   {r \over n-1} \, {(r-1)n \choose n-2}
               \;=\;   r \, {[(r-1)n]!   \over  (n-1)! \, [(r-2)n + 2]!}
      \label{tr_result}  \\[3mm]
   u_n^{(r)}   & = &   {1 \over n} \, {(r-1)n \choose n-1}
               \;=\;   (r-1) \, {[(r-1)n -1]!  \over  (n-1)! \, [(r-2)n + 1]!}
      \label{ur_result}
\end{eqnarray}
\qed

\medskip
\par\noindent
{\bf Remarks.}  1.  The proof presented here is a simplification
of my original proof, based on independent suggestions by
Paul Seymour, Dave Wagner and an anonymous referee.

2. {\em A posteriori}\/, we learn from Proposition \ref{prop4.3}(c,d) below
that the power series \reff{Tr_def}/\reff{Ur_def} in fact define
analytic functions in the disc $|z| < (r-2)^{r-2}/(r-1)^{r-1}$.

3.  I suspect that Proposition \ref{prop4.2} is known somewhere
in the graph-theory literature, but I do not know any reference.
A weaker version of Proposition \ref{prop4.2}
can be found in \cite[Lemma 5.4]{Dobrushin_96b}.

\bigskip

Let us also collect some properties of the numbers $t_n^{(r)}$
that arise in Proposition \ref{prop4.2}:

\begin{proposition}
   \label{prop4.3}
The quantities
\be
   t_n^{(r)}   \;=\;  r \, {[(r-1)n]!   \over  (n-1)! \, [(r-2)n + 2]!}   \;,
\ee
defined for integers $n,r \ge 1$, have the following properties:
\begin{itemize}
   \item[(a)]  $t_n^{(1)} = \cases{1  & for $n=1,2$ \cr
                                   0  & for $n \ge 3$ \cr
                                  }$
   \item[(b)]  $t_n^{(2)} = n$
   \item[(c)]  As $n \to \infty$ at fixed $r \ge 3$,
\be
   t_n^{(r)}   \;=\;
      {r \, (r-1)^{1/2}  \over \sqrt{2\pi} \, (r-2)^{5/2}}
      \left( {(r-1)^{r-1} \over (r-2)^{r-2}} \right) ^{\! n}  \, n^{-3/2}
   \left[ 1 \,+\, {{1 \over r-1} - {37 \over r-2} - 1  \over 12n}
                               \,+\, O\!\left( {1 \over n^2} \right) \right]   \;.
\ee
   \item[(d)]  For all $n$ and all $r \ge 3$,
\be
   t_n^{(r)}   \;\le\;  \left( {(r-1)^{r-1} \over (r-2)^{r-2}} \right) ^{\! n-1}
               \;\le\;  [e(r-{\textstyle{3 \over 2}})]^{n-1}   \;.
\ee
   \item[(e)]  As $r \to \infty$ at fixed $n \ge 1$,
\be
   t_n^{(r)}   \;=\;  {(rn)^{n-1} \over n!}
      \left[ 1 \,-\, {3(n-1)(n-2) \over 2nr}
               \,+\, O\!\left( {1 \over r^2} \right) \right]
   \;.
\ee
   \item[(f)]  For all $r,n \ge 1$,
\be
   t_n^{(r)}   \;\le\;  {(rn)^{n-1} \over n!}
   \;.
\ee
\end{itemize}
\end{proposition}

\proof
(a) and (b) are trivial,
while (c) and (e) follow from Stirling's formula.
(f) is trivial for $r=1$, while for $r \ge 2$ it follows immediately from
\be
   {[(r-1)n]! \over [(r-2)n + 2]!}   \;=\;
   {(rn-n)! \over (rn-2n+2)!}   \;\le\;
   (rn)^{n-2}   \;.
\ee
The first inequality in (d) is obvious for $n=1$, so assume $n \ge 2$.
We have
\begin{eqnarray}
   t_n^{(r)}
      & = &   {rn \over [(r-2)n + 1] [(r-2)n + 2]} \,  {(r-1)n \choose n}
   \nonumber \\[1mm]
      &\le&   {r \over (r-2)^2 n} \,  {(r-1)n \choose n}
   \nonumber \\[1mm]
      &\le&   {r (r-1)^{1/2}  \over \sqrt{2\pi} (r-2)^{5/2}} \, n^{-3/2}
                    \left( {(r-1)^{r-1} \over (r-2)^{r-2}} \right) ^{\! n}
   \nonumber \\[1mm]
      & = &   {r (r-1)^{r - {1 \over 2}}  \over
               \sqrt{2\pi} (r-2)^{r + {1 \over 2}}} \, n^{-3/2}
                    \left( {(r-1)^{r-1} \over (r-2)^{r-2}} \right) ^{\! n-1}
              \;,
\end{eqnarray}
where the second inequality uses Lemma \ref{lemma4.4} below.
Then straightforward calculus shows that the function
$F(r) = r (r-1)^{r - {1 \over 2}} / [\sqrt{2\pi} (r-2)^{r + {1 \over 2}}]$
is decreasing on $r>2$;
%
%
and we have
$F(3) = 12/\sqrt{\pi} < 4^{3/2}$,
$F(4) = (27/8)  \sqrt{3/\pi} < 3^{3/2}$
and $F(5) = (1280/243) \sqrt{2/(3\pi)} < 2^{3/2}$.
So the first inequality in (d) follows except for the cases
$(r,n) = (3,2), \, (3,3)$ and $(4,2)$,
which can be checked by hand.
The final inequality in (d) follows from
\begin{eqnarray}
   \log  {\sigma^\sigma  \over (\sigma-1)^{\sigma-1}}
   & = &   \log\sigma \,+\, (\sigma-1) \log {\sigma  \over \sigma-1}
      \nonumber \\[2mm]
   & = &   \log\sigma \,+\, 1  \,-\,
               \sum\limits_{k=1}^\infty  {\sigma^{-k} \over k(k+1)}
      \nonumber \\[2mm]
   &\le&   \log\sigma \,+\, 1  \,-\,
               \sum\limits_{k=1}^\infty  {\sigma^{-k} \over k 2^k}
      \nonumber \\[2mm]
   & = &   \log\sigma \,+\, 1  \,+\, \log\!\left(1 - {1 \over 2\sigma}\right)
\end{eqnarray}
where the sums are convergent for $\sigma > 1$,
so that $\sigma^\sigma / (\sigma-1)^{\sigma-1} \le e(\sigma - {1 \over 2})$.
\qed

\begin{lemma}
   \label{lemma4.4}
Let $n \ge 2$ and $1 \le k \le n-1$ be integers.  Then
\be
   {n \choose k}   \;<\;
   \left( {n \over k} \right)  ^{\! k}
   \left( {n \over n-k} \right)  ^{\! n-k}
   \sqrt{ {n \over 2\pi k (n-k)} }
   \;.
\ee
\end{lemma}

\proof
We use the following strong form of Stirling's formula
\cite[pp.~45--46]{Chow_97}:
for integer $n \ge 1$,
\be
   \log n!   \;=\;  (n + \smhalf) \log n  \,-\, n \,+\, \log \sqrt{2\pi}
                      \,+\, \epsilon_n
\ee
with
\be
   {1 \over 12n+1}  \;<\;  \epsilon_n   \;<\;   {1 \over 12n}   \;\;.
\ee
(The proof in \cite{Chow_97} is valid only for $n \ge 2$,
 but $\epsilon_1 = 1 - \log \sqrt{2\pi} \approx 0.08106$
 clearly satisfies $1/13 < \epsilon_1 < 1/12$.)
Then
\begin{eqnarray}
   \epsilon_n - \epsilon_k - \epsilon_{n-k}
      & < &  {1 \over 12n} \,-\, {1 \over 12k+1}  \,-\,  {1 \over 12(n-k) + 1}
   \nonumber \\[1mm]
      & = &  {-144[n^2 - k(n-k)] - 12n + 1
              \over
              12n \, (12k+1) \, [12(n-k) + 1]
             }
   \nonumber \\[1mm]
       & < &  0   \;.
\end{eqnarray}
\qed

Proposition \ref{prop4.2} clearly gives the best possible bound
for $C_{\bullet,m}(G, x)$ and $T_n(G, x)$
in terms of the maximum degree of $G$,
since it is sharp when $G = {\bf T}_r$.
On the other hand, Proposition \ref{prop4.2}
is somewhat unnatural for general (unequal) edge weights $\{v_e\}$,
since adding an edge of small weight $v_e$ makes little change in
$C_{n,m}(G, \{ v_e \}, x)$ but can cause the bound to jump
(in case it increases the maximum degree).
It is of interest, therefore, to find alternative bounds that
depend ``smoothly'' on the weights $\{v_e\}$.
We shall now give two such bounds
(Propositions \ref{prop4.5} and \ref{prop4.6}).
Unfortunately, both of them are strictly weaker than Proposition \ref{prop4.2}
when the edge weights are equal,
and neither one is strictly stronger than the other.

\begin{proposition}
   \label{prop4.5}
Let $G=(V,E)$ be a finite or countably infinite loopless undirected graph
equipped with edge weights $\{ v_e \}_{e \in E}$.
Then for any $x \in V$,
\begin{subeqnarray}
   C_{\bullet,m}(G, \{ v_e \}, x)   & \le &
   {(m+1)^m \over (m+1)!} \,
   \left( \sup\limits_{i \in V} \sum\limits_{e \ni i} |v_e| \right) ^{\! m}
       \\[2mm]
   & \le &
   \left( e \, \sup\limits_{i \in V} \sum\limits_{e \ni i} |v_e| \right) ^{\! m}
   \;.
       \slabel{eq.prop4.5.b}
\end{subeqnarray}
[The $e$ in front of the sup in \reff{eq.prop4.5.b} denotes, of course,
 the base of the natural logarithms!]
\end{proposition}

\proof
As in the proof of Proposition \ref{prop4.2},
we pass to the universal covering graph $U = (\widetilde{V}, \widetilde{E})$
of $G$ with covering map $f \colon\, U \to G$;
and we define the weight $v_{\widetilde{e}}$
of an edge $\widetilde{e} \in \widetilde{E}$
to be the weight $v_e$ of its image $e = f(\widetilde{e})$.
It then follows, as in Proposition \ref{prop4.2}, that
$C_{\bullet,m}(G, \{v_e\}, x)  \le
 C_{\bullet,m}(U, \{v_{\widetilde{e}}\}, \widetilde{x})$.

Let us now define, for each vertex $x \in \widetilde{V}$,
the formal generating function
\be
   C_x(z)  \;=\;  \sum\limits_{m=0}^\infty
                  C_{\bullet,m}(U, \{v_{\widetilde{e}}\}, x) \, z^m
   \;.
\ee
Then the recursive structure of rooted trees implies that
\begin{subeqnarray}
   C_x(z)  & \preceq &  \prod\limits_{y \sim x} [1 + |v_{xy}| z C_y(z)]   \\[2mm]
           & \preceq &  \prod\limits_{y \sim x} e^{|v_{xy}| z C_y(z)}
\end{subeqnarray}
where $y \sim x$ denotes that $y$ is adjacent to $x$,
$xy$ denotes the (unique) corresponding edge,
and $\preceq$ denotes coefficientwise inequality at all orders in $z$;
the second inequality holds because
$1 + \alpha z \preceq e^{\alpha z}$ for $\alpha \ge 0$.
It then follows, by induction on the power of $z$,
that $C_x(z) \preceq \bar{C}(z)$ for all $x$,
where $\bar{C}(z)$ is determined by the equation
\be
   \bar{C}(z)   \;=\;  e^{\mu z \bar{C}(z)}
\ee
with
\be
   \mu   \;=\;
   \sup\limits_{x \in \widetilde{V}} \sum\limits_{y \sim x} |v_{xy}|
   \;=\;
   \sup\limits_{i \in V} \sum\limits_{e \ni i} |v_e|
   \;.
\ee
The coefficients of $\bar{C}(z)$ can be determined by applying the
Lagrange Implicit Function Theorem to $\bar{U}(z) = z \bar{C}(z)$,
and we have the well-known (e.g. \cite[p.~392]{Knuth_73}) result
\be
   \bar{C}(z)   \;=\;
   \sum_{m=0}^\infty   {(m+1)^m \over (m+1)!} \mu^m z^m
   \;.
\ee
Finally, the inequality $e^m \ge (m+1)^m / (m+1)!$ is trivial for $m=0,1$,
and for $m \ge 2$ it follows from $e^x \ge x^n/n!$ by setting $x = n = m+1$.
\qed

\medskip
\par\noindent
{\bf Remarks.}  1.  The proof presented here is a simplification
of my original proof, based on the suggestions of an anonymous referee.

2. Proposition \ref{prop4.5} holds also for graphs with loops,
but they impose slight technical complications.

\bigskip

An alternative estimate is due to
Campanino {\em et al.}\/ \cite[p.~129]{Campanino_79}
(see also \cite[p.~522]{Cammarota_82} and \cite[pp.~463--464]{Simon_93}):

\begin{proposition}
   \label{prop4.6}
Let $G=(V,E)$ be a finite or countably infinite undirected graph
equipped with edge weights $\{ v_e \}_{e \in E}$.
Define the matrix $M = (M_{xy})_{x,y \in V}$ by
\be
   M_{xy}   \;=\;
   \sum_{e\colon\; \hboxscript{$e$ connects $x$ to $y$}} \!  |v_e|^{1/2}
   \;.
\ee
Then for any $x \in V$,
\be
   C_{\bullet,m}(G, \{ v_e \}, x)  \;\le\;
   (M^{2m})_{xx}   \;\le\;
   \left( \sup\limits_{i \in V} \sum\limits_{e \ni i}
                                         |v_e|^{1/2} \right) ^{\! 2m}
   \;.
\ee
\end{proposition}

\proof
Let $G' = (V',E')$ be a connected subgraph of $G$ having $m$ edges.
Then for any vertex $x \in V'$,
there exists a path on $G'$ starting and ending at $x$
that uses each edge $e \in E'$ exactly twice.
(Proof: The multigraph formed by doubling each edge of $G'$ is Eulerian.
 Alternate proof: By induction on $m$.\footnote{
    See e.g.\ \cite[Lemma V.7.A.2]{Simon_93}.
})
Conversely, every path on $G$ starting and ending at $x$
corresponds in this way to at most one subgraph $G'$.
The claim follows.
\qed

\bigskip
Let us conclude by examining the relative sharpness of these bounds
when $G$ is an $r$-regular graph and the edge weights $v_e$ are equal.
Then the tree bound $t_n^{(r)}$ of Proposition \ref{prop4.2}
grows as $n\to\infty$ at an exponential rate $(r-1)^{r-1} / (r-2)^{r-2}$:
this is less than $e(r-{3 \over 2})$ for all $r$,
and behaves as $e [ r - {3 \over 2} - O(1/r)]$ as $r \to\infty$.
The bound of Proposition \ref{prop4.5} is slightly weaker:
it grows at exponential rate $er$.
Finally, the bound of Proposition \ref{prop4.6}
grows at exponential rate $r^2$,
which is vastly weaker for large $r$ but is slightly better when $r=2$.

In particular, when $G$ is a regular lattice, it can be shown by
supermultiplicativity arguments
\cite{Klarner_67,Klein_81,Whittington_90,Janse_92}
that the limits
\begin{eqnarray}
   \lambda_o(G)   & = &
        \lim\limits_{n\to\infty}   T_n(G,x)^{1/n}
     \\[2mm]
   \lambda_b(G)   & = &
        \lim\limits_{m\to\infty}   C_{\bullet,m}(G,x)^{1/m}
\end{eqnarray}
exist.
For the simple hypercubic lattice $\Z^d$ with nearest-neighbor bonds,
these growth constants have been computed (non-rigorously)
in a large-$d$ asymptotic expansion \cite{Gaunt_94,Peard_95}
(see also \cite{Penrose_94,Hara_95,Derbez_98} for related rigorous results):
\begin{eqnarray}
   \log\lambda_o(\Z^d)   & = &
        \log\sigma \,+\, 1 \,-\, {1 \over 2} \sigma^{-1}
            \,-\, {8 \over 3} \sigma^{-2}
            \,-\, \ldots
    \\[3mm]
   \log\lambda_b(\Z^d)   & = &
        \log\sigma \,+\, 1 \,-\, {1 \over 2} \sigma^{-1}
            \,- \left( {8 \over 3} - {1 \over 2e} \right) \sigma^{-2}
            \,-\, \ldots
\end{eqnarray}
where $\sigma = r-1 = 2d-1$.
Let us compare this with the tree bound of Proposition \ref{prop4.2}:
\be
   \log {(r-1)^{r-1} \over (r-2)^{r-2}}
   \;=\;
   \log\sigma \,+\, 1 \,-\, {1 \over 2} \sigma^{-1}
            \,-\, {1 \over 6} \sigma^{-2}
            \,-\, \ldots
   \;.
\ee
Thus, the latter bound is very close to sharp for $G = \Z^d$
in high dimension $d$, confirming the intuition that
high-dimensional regular lattices are ``like trees'' to leading order in $1/d$.

\section{Application to the Potts-Model Partition Function}   \label{sec5}

We are now ready for the main theorem of this paper:

\begin{theorem}
   \label{thm5.1}
Let $G=(V,E)$ be a loopless finite undirected graph
equipped with complex edge weights $\{ v_e \}_{e \in E}$
satisfying $|1 + v_e| \le 1$ for all $e$.
Let $Q = Q(G, \{ v_e \}) > 0$ be the smallest number for which
\be
   \inf\limits_{\alpha > 0}
   \alpha^{-1} \sum\limits_{n=2}^\infty  e^{\alpha n} \, Q^{-(n-1)} \,
      \max\limits_{x \in V} T_n(G, \{ v_e \}, x)
   \;\,\le\;\,   1
   \;.
 \label{eq5.1}
\ee
[Note that $Q$ is automatically finite,
 since $T_n(G, \{ v_e \}, x) = 0$ for $n > |V|$.]
Then all the zeros of $Z_G(q, \{v_e\})$ lie in the disc $|q| < Q$.
\end{theorem}

\proof
Starting from the polymer-gas representation \reff{eq2.2}--\reff{eq2.4}
of $Z_G(q, \{v_e\})$,
we apply Theorem \ref{thm3.1} and Proposition \ref{prop3.2}
with the choice $R_S = |w(S)|$.
We verify hypothesis (a) of Proposition \ref{prop3.2}
by verifying \reff{partition_condition_2} with
\be
   A_n   \;=\;
   \max_{x \in V}  \sum_{\begin{scarray}
                            S \ni x  \\
                            |S| = n
                         \end{scarray}} 
                   |w(S)|
   \;.
 \label{eq5.2}
\ee
Now we use Proposition \ref{prop_penrose}
to conclude that $w(S)$ can be bounded by a sum over trees:
\be
   |w(S)|   \;\le\;  |q|^{-(|S|-1)}
           \sum\limits_{\begin{scarray}
                            \widetilde{E} \subseteq E  \\
                            (S,\widetilde{E}) \hboxscript{ tree}
                        \end{scarray}}
           \; \prod\limits_{e \in \widetilde{E}} |v_e|   \;.
\ee
Inserting this into \reff{eq5.2}, we get
\be
   A_n  \;\le\;  |q|^{-(n-1)} \, \max_{x \in V}  T_n(G, \{ v_e \}, x)   \;.
\ee
If $|q| \ge Q$, hypothesis (b) of Proposition \ref{prop3.2} holds
(recall Remark 4 following that Proposition)
and hence $Z_G(q, \{v_e\}) \neq 0$.
\qed

In applying Theorem \ref{thm5.1} we are of course free to use
any convenient upper bound on $\max_{x \in V} T_n(G, \{ v_e \}, x)$.
In particular, when $G$ has maximum degree $\le r$,
Proposition \ref{prop4.2} provides such a bound.
Recall that
\be
   t_n^{(r)}  \;=\;  r \, {[(r-1)n]!   \over  (n-1)! \, [(r-2)n + 2]!}   \;,
\ee
and let $C = C(r) > 0$ be the smallest number for which
\be
   \inf\limits_{\alpha > 0}
   \alpha^{-1} \sum\limits_{n=2}^\infty  e^{\alpha n} \, C^{-(n-1)} \, t_n^{(r)}
   \;\le\;   1
   \;.
 \label{Falpha}
\ee
The following is then an immediate consequence of Theorem \ref{thm5.1}
and Proposition \ref{prop4.2}:

\begin{corollary}
\label{cor5.2}
Let $G=(V,E)$ be a loopless finite undirected graph
of maximum degree $\le r$,
equipped with complex edge weights $\{ v_e \}_{e \in E}$
satisfying $|1 + v_e| \le 1$ for all $e$.
Let $v_{max} = \max\limits_{e \in E} |v_e|$.
Then all the zeros of $Z_G(q, \{v_e\})$ lie in the disc $|q| < C(r) v_{max}$.
\end{corollary}

\noindent
And for the chromatic polynomials:

\begin{corollary}
\label{cor5.3}
Let $G=(V,E)$ be a loopless finite undirected graph
of maximum degree $\le r$.
Then all the zeros of $P_G(q)$ lie in the disc $|q| < C(r)$.
\end{corollary}

Table \ref{table1} lists rigorous upper bounds on $C(r)$ for $2 \le r \le 20$,
proven (with the assistance of {\sc Mathematica}) as follows:
After computing numerically an approximate value of $C(r)$,\footnote{
    Using the generating function $f(z) = \sum_{n=2}^\infty t_n^{(r)} z^n$
    where $z = e^{\alpha}/C$, I solve simultaneously the equations
    $f(z)  = (\log z + \log C) / C$ and $f'(z) = 1/(Cz)$
    by solving numerically $f(z) = - z f'(z) \log f'(z)$
    and then plugging back in to determine
    $C = 1/[z f'(z)]$ and $\alpha = - \log f'(z)$.
}
I added $10^{-6}$ and rounded it upwards to a rational number $p/10^6$.
(Thus, the value reported in Table \ref{table1}
 exceeds my best estimate of $C(r)$ by at most $2 \times 10^{-6}$.)
I likewise approximated the numerically-found $\alpha$
by a rational number $p'/10^6$.
Thereafter I did all computations in exact rational arithmetic.
First I computed a rational upper bound on $e^\alpha$
(differing from the true $e^{\alpha}$ by at most $2 \times 10^{-10}$)
by truncating the Taylor series for $e^{-\alpha}$ at odd order
(here ninth or eleventh)
to obtain a lower bound on $e^{-\alpha}$.
Finally, I computed an upper bound on \reff{Falpha}
by summing the terms explicitly through $n=$ some $n_0$
and bounding the tail of the series ($n \ge n_0 + 1$)
using Proposition \ref{prop4.3}(d);
I systematically increased $n_0$ until
the inequality \reff{Falpha} was verified.
For $r=2$, of course, I just summed the series exactly.

As $r \to \infty$ we have the following:

\begin{table}
\begin{center}
\begin{tabular}{rrrrc}
\multicolumn{1}{c}{$r$}  &
   \multicolumn{1}{c}{$C(r)$} &
   \multicolumn{1}{c}{$\alpha$} &
   \multicolumn{1}{c}{$e^{\alpha}$} &
   \multicolumn{1}{c}{$n_0$}   \\
\hline \\[-4mm]
  2 &   13.234367 &  0.453177 &  1.5733026366 & $\infty$ \\
  3 &   21.144294 &  0.436884 &  1.5478765131 & 15 \\
  4 &   29.081607 &  0.428653 &  1.5351882318 & 18 \\
  5 &   37.029702 &  0.423694 &  1.5275940786 & 20 \\
  6 &   44.983130 &  0.420382 &  1.5225430561 & 22 \\
  7 &   52.939585 &  0.418013 &  1.5189404206 & 23 \\
  8 &   60.897921 &  0.416236 &  1.5162436603 & 24 \\
  9 &   68.857505 &  0.414852 &  1.5141466306 & 25 \\
 10 &   76.817961 &  0.413745 &  1.5124713976 & 25 \\
 11 &   84.779049 &  0.412839 &  1.5111017191 & 26 \\
 12 &   92.740610 &  0.412084 &  1.5099612679 & 26 \\
 13 &  100.702534 &  0.411445 &  1.5089967109 & 26 \\
 14 &  108.664743 &  0.410898 &  1.5081715154 & 27 \\
 15 &  116.627179 &  0.410423 &  1.5074553040 & 27 \\
 16 &  124.589800 &  0.410008 &  1.5068298399 & 28 \\
 17 &  132.552573 &  0.409641 &  1.5062769348 & 28 \\
 18 &  140.515473 &  0.409315 &  1.5057859685 & 28 \\
 19 &  148.478479 &  0.409024 &  1.5053478486 & 28 \\
 20 &  156.441575 &  0.408761 &  1.5049519941 & 28 
\end{tabular}
\end{center}
\caption{
   Upper bounds on $C(r)$ for $2 \le r \le 20$;
   they differ from my best estimate of the true $C(r)$
   by at most $2 \times 10^{-6}$.
   The third column gives a value of $\alpha$ for which (\protect\ref{Falpha})
   is proven to be $\le 1$.
   The fourth column gives the upper bound on $e^{\alpha}$
   employed in this proof.
   The fifth column gives the number of terms explicitly summed in the series.
}
\label{table1}
\end{table}

\begin{proposition}
   \label{prop5.4}
Let $K \approx 7.963906\ldots$ be the smallest number for which
\be
   \inf\limits_{\alpha > 0}
   \alpha^{-1} \sum\limits_{n=2}^\infty  e^{\alpha n} \, K^{-(n-1)} \,
       {n^{n-1} \over n!}
   \;\le\;   1
   \;.
\ee
Then $C(r) \le Kr$ for all $r$,
and $\lim\limits_{r \to\infty} C(r)/r = K$.
Moreover, we have the rigorous bound $K \le 7.963907$.
\end{proposition}

\proof
Clearly $\widetilde{C}(r) \equiv C(r)/r$ is the smallest number for which
\be
   \inf\limits_{\alpha > 0}
   \alpha^{-1} \sum\limits_{n=2}^\infty  e^{\alpha n} \,
       \widetilde{C}^{-(n-1)} \, {t_n^{(r)} \over r^{n-1}}
   \;\le\;   1
   \;.
\ee
It follows from Proposition \ref{prop4.3}(f) that
$\widetilde{C}(r) \le K$ for all $r$.

Now suppose that we were to have
$\liminf\limits_{r \to \infty} \widetilde{C}(r) \le K-\epsilon < K$.
Then there would exist infinite sequences $\{ r_i \} \uparrow \infty$
and $\{ \alpha_i \}$ such that
\be
   \alpha_i^{-1} \sum\limits_{n=2}^\infty  e^{\alpha_i n} \,
       (K-\epsilon)^{-(n-1)} \, {t_n^{(r_i)} \over r_i^{n-1}}
   \;\le\;   1
  \label{prop5.4_star1}
\ee
for all $i$.  Now the finiteness of \reff{prop5.4_star1} implies
that the $\alpha_i$ are bounded
[e.g.\ from Proposition \ref{prop4.3}(c) we have
 $e^{\alpha_i} \le (K-\epsilon) r_i (r_i -2)^{r_i -2} / (r_i -1)^{r_i -1}
               \le {3 \over 4} (K-\epsilon)$
 whenever $r_i \ge 3$].
So we can extract a subsequence of $\{ \alpha_i \}$ that converges to
some value $\alpha_*$.  Then Proposition \ref{prop4.3}(e,f) and
the dominated convergence theorem imply that
\be
   \alpha_*^{-1} \sum\limits_{n=2}^\infty  e^{\alpha_* n} \,
       (K-\epsilon)^{-(n-1)} \, {n^{n-1} \over n!}
   \;\le\;   1
   \;,
\ee
which contradicts the definition of $K$.

Finally, it is easy to prove that $K \le 7.963907$,
by a computer-assisted method similar to that used above for $C(r)$.
For the tail of the series ($n \ge n_0 + 1$),
it suffices to use the crude bound
$n^{n-1}/n! \le e^{n-1} \le 3^{n-1}$.
The proof succeeds with the choices
$\alpha = 0.403774$, $e^\alpha \le 1.4974655$ and $n_0 = 32$.
\qed

If we employ Proposition \ref{prop4.5} in place of Proposition \ref{prop4.2},
Theorem \ref{thm5.1} yields the following:

\begin{corollary}
   \label{cor5.5}
Let $G=(V,E)$ be a loopless finite undirected graph
equipped with complex edge weights $\{ v_e \}_{e \in E}$
satisfying $|1 + v_e| \le 1$ for all $e$.
Then all the zeros of $Z_G(q, \{v_e\})$ lie in the disc
$|q| < K \max\limits_{i \in V} \sum\limits_{e \ni i} |v_e|$,
where $K$ is the constant defined in Proposition \ref{prop5.4}.
\end{corollary}

\medskip
\par\noindent
{\bf Remarks.}  1.  Since $Z_G(q, \{v_e\})/q$ for any graph $G$ is the product
of the same quantity over the blocks of $G$, it is legitimate to apply
Theorem \ref{thm5.1} and its corollaries separately to each block.
This can lead to large improvements (consider e.g.\ trees).

2.  What happens if we drop the assumption that $|1 + v_e| \le 1$?
Because we can no longer use Proposition \ref{prop_penrose}
to reduce the sum to trees,
we need to consider all $n$-vertex connected subgraphs of $G$
containing a given vertex $x$.
But the number $m$ of {\em edges}\/ in such a subgraph
could be as large as $\lfloor rn/2 \rfloor$
(where $r$ is the maximum degree of $G$).
Therefore, the factor $t_n^{(r)} v_{max}^{n-1}$
coming from \reff{eq4.17}
has to be replaced by a factor
$\sum_{m=n-1}^{\lfloor rn/2 \rfloor} t_{m+1}^{(r)} v_{max}^m$
coming from \reff{eq4.16}
[or the analogue from Proposition \ref{prop4.5}].
As a consequence, the radius of the $q$-plane disc
containing all the zeros of $Z_G(q, \{v_e\})$
will scale as $\max[v_{max}, v_{max}^{r/2}]$ rather than simply $v_{max}$.
And this is not simply an artifact of the method of proof:
for the $q$-state Potts {\em ferromagnet}\/ ($v > 0$, $q > 0$)
on the the simple hypercubic lattice $\Z^d$ with nearest-neighbor bonds,
the first-order phase-transition point $v_t$ indeed behaves as
\be
   v_t(q)   \;=\;  q^{1/d} \, [1 + O(1/q)]
\ee
as $q \to +\infty$
\cite{Martirosian_86,Laanait_86,Kotecky_90,Laanait_91,Borgs_91}.
In the Whitney--Tutte--Fortuin--Kasteleyn polynomial \reff{eq1.2},
this reflects the coexistence at $v=v_t$ (for all $q \gg 1$)
between a phase with a low density of occupied edges
and a phase with a high density of occupied edges.

\section{Some Generalizations}   \label{sec6}

The following generalization of the
Whitney--Tutte--Fortuin--Kasteleyn polynomial \reff{eq1.2}
is motivated by some work of
Tutte \cite{Tutte_47,Tutte_76}, Farrell \cite{Farrell_79}
and Stanley \cite{Stanley_95,Stanley_98}
as well as by the statistical-mechanical application to be discussed below.
Let us replace the single complex number $q$
by a map $\q\colon\, \scrp_*(V) \to \C$, and define
\be
   Z_G(\q, \{v_e\})   \;=\;
   \sum_{ E' \subseteq E }  \left( \prod_{i=1}^{k(E')} \q(V_i) \right)
                          \left( \prod_{e \in E'}  v_e \right)
   \;,
 \label{eq6.1}
\ee
where $(V_1,E_1), \ldots, (V_{k(E')}, E_{k(E')})$
are the connected components of $(V,E')$.
We immediately deduce an analogue of Proposition \ref{prop2.1}:
the identity \reff{eq2.2} is replaced by
\be
   Z_G(\q, \{v_e\})   \;=\;
      \left( \prod\limits_{x \in V} \q(\{x\}) \right)
      Z_{polymer,G}(\q, \{v_e\})
   \;,
   \label{eq6.2}
\ee
and the fugacities $w(S)$ are now given by
\be
   w(S)   \;=\;
   \cases{ {\displaystyle \q(S)
            \over
            \displaystyle \prod\limits_{x \in S} \q(\{x\})
           }  \!\!
           \sum\limits_{\begin{scarray}
                            \widetilde{E} \subseteq E  \\
                            (S,\widetilde{E}) \hboxscript{ connected}
                        \end{scarray}}
           \prod\limits_{e \in \widetilde{E}} v_e
           & if $|S| \ge 2$   \cr
           \noalign{\vskip 2mm}
           0   &  if $|S| \le 1$ \cr
         }
  \label{eq6.3}
\ee
The proof of Theorem \ref{thm5.1} then goes through without change,
and yields:

\begin{theorem}
   \label{thm6.1}
Let $G=(V,E)$ be a loopless finite undirected graph
equipped with complex edge weights $\{ v_e \}_{e \in E}$
satisfying $|1 + v_e| \le 1$ for all $e$,
and let $\q\colon\, \scrp_*(V) \to \C$.
Let $\{ Q_n \} _{n=1}^\infty$ be a sequence of positive numbers satisfying
\be
   \inf\limits_{\alpha > 0}
   \alpha^{-1} \sum\limits_{n=2}^\infty  e^{\alpha n} \, Q_n^{-1} \,
      \max\limits_{x \in V} T_n(G, \{ v_e \}, x)
   \;\,\le\;\,   1
   \;,
 \label{eq6.4}
\ee
and assume that
\be
   \left|  {\q(S) \over \prod\limits_{x \in S} \q(\{x\})}  \right|
   \;\le\;  Q_{|S|}^{-1}
 \label{eq6.5}
\ee
for all nonempty subsets $S \subseteq V$.
Then $Z_G(\q, \{v_e\}) \neq 0$.
\end{theorem}

The following special case is of particular interest:
Fix an integer $N \ge 0$, and for each $x \in V$
choose a vector $u_x = (u_x^{(1)}, \ldots, u_x^{(N)}) \in \C^N$.
Then define
\be
   \q(S)   \;=\;  q \,-\, N \,+\, \sum_{i=1}^N \prod_{x \in S} (1+ u_x^{(i)})
 \label{eq6.star1}
\ee
where $q$ is a fixed complex number.
This corresponds to a $q$-state Potts model
in a magnetic field $h_x = (h_x^{(1)}, \ldots, h_x^{(N)})$
in the first $N$ spin directions, where $u_x^{(i)} = \exp(h_x^{(i)}) - 1$.
To see this, we first define, for each integer $q \ge N$,
the partition function for the $q$-state Potts model in a magnetic field,
generalizing \reff{eq1.1}:
\be
   Z_G(q, \{v_e\}, \{u_x^{(i)}\})   \;=\;
   \sum_{ \{\sigma_x\} }  \,
   \prod_{e \in E}  \,
      \biggl[ 1 + v_e \delta(\sigma_{x_1(e)}, \sigma_{x_2(e)}) \biggr]  \,
   \prod_{x \in V} \prod_{i=1}^N \,
         \biggl[ 1 + u_x^{(i)} \delta(\sigma_x, i) \biggr]
   \;.
 \label{eq6.star2}
\ee
Now expand out the product over $e \in E$,
and let $E' \subseteq E$ be the set of edges for which the term
$v_e \delta_{\sigma_{x_1(e)}, \sigma_{x_2(e)}}$ is taken.
Then perform the sum over configurations $\{ \sigma_x \}$:
in each connected component of the subgraph $(V,E')$
the spin value $\sigma_x$ must be constant,
and there are no other constraints.
The sum over possible spin values in a connected component
with vertex set $S$ yields \reff{eq6.star1}.
It follows that, for any integer $q \ge N$,
the partition function $Z_G(q, \{v_e\}, \{u_x^{(i)}\})$
equals $Z_G(\q, \{v_e\})$ with weights \reff{eq6.star1}.
We then take the latter, which is a polynomial in $q$, $\{v_e\}$
and $\{u_x^{(i)}\}$, as the {\em definition}\/ of
$Z_G(q, \{v_e\}, \{u_x^{(i)}\})$ for general complex $q$.

The following lemma gives a sufficient condition for the
applicability of Theorem \ref{thm6.1} to this situation:

\begin{lemma}
   \label{lemma6.2}
Let $\q(S)$ be defined by \reff{eq6.star1}.
\begin{itemize}
   \item[(a)]  If $N=1$ and each $u_x^{(i)}$ equals either $0$ or $-1$, then
\be
   \left|  {\q(S) \over \prod\limits_{x \in S} \q(\{x\})}  \right|
   \;\le\;  \min(|q|, |q-1|)^{-(|S|-1)}   \;.
 \label{eq_lemma6.2a}
\ee
%
   \item[(b)]  If $-1 \le u_x^{(i)} \le 0$ for all $x,i$ and $|q| > N$, then
\be
   \left|  {\q(S) \over \prod\limits_{x \in S} \q(\{x\})}  \right|
   \;\le\;  (|q| - N)^{-(|S|-1)}   \;.
 \label{eq_lemma6.2b}
\ee
%
   \item[(c)]  If $|1 + u_x^{(i)}| \le 1$ for all $x,i$
and $|q-N| > N$, then
\be
   \left|  {\q(S) \over \prod\limits_{x \in S} \q(\{x\})}  \right|
   \;\le\;  {|q-N| + N   \over (|q-N|-N)^{|S|} }   \;.
 \label{eq_lemma6.2c}
\ee
\end{itemize}
\end{lemma}

\noindent
The proof of Lemma \ref{lemma6.2} is deferred to the end of this section.

We can exploit this example to obtain new results for the
ordinary (zero-field) Potts-model partition function $Z_G(q,\{v_e\})$
and in particular for the chromatic polynomial $P_G(q)$,
by employing a variant of the ``ghost spin'' trick
of Suzuki \cite{Suzuki_65} and Griffiths \cite{Griffiths_67}.
Given a finite graph $G_0 = (V_0, E_0)$ and an integer $N \ge 1$,
we define $G$ to be the join of $G_0$ with the complete graph on $N$ vertices.
Thus, the vertex set of $G$ is
$V = V_0 \bigcup \{y_1,\ldots,y_N\}$  (disjoint union)
and the edge set is
$E = E_0 \bigcup \{ \<x y_i\> \}_{x \in V_0, \, 1 \le i \le N}
         \bigcup \{ \<y_i y_j\> \}_{ 1 \le i < j \le N}$.
We allow the edge weights
$\{ v_e \} _{e \in E_0}$ and
$\{ v_{\<x y_i\>} \} _{x \in V_0, \, 1 \le i \le N}$
to be arbitrary complex numbers,
but we require that $v_{\<y_i y_j\>} = -1$ for $1 \le i < j \le N$
(this condition is crucial).
We then have the identity
\be
   Z_G(q, \{v_e, v_{\<x y_i\>}, v_{\<y_i y_j\>} \})   \;=\;
      q_{\<N\>} \, Z_{G_0}(q, \{v_e\}, \{u_x^{(i)}\})
   \;,
 \label{eq6.star3}
\ee
where $q_{\<N\>} = q (q-1) \cdots (q-N+1)$
is the $N$-th ``falling factorial'' polynomial
and $u_x^{(i)} = v_{\<x y_i\>}$.
This is most easily proven in the Potts spin representation
\reff{eq1.1}/\reff{eq6.star2}:
Let $q$ be an integer $\ge N$,
and let us compute the left-hand side of \reff{eq6.star3}.
There are $q_{\<N\>}$ admissible ways
to color the vertices $\{y_1,\ldots,y_N\}$,
all of which are equivalent modulo permutations of $\{1,\ldots,q\}$;
and with any such coloring fixed, the sum over colorings of $V_0$
yields precisely $Z_{G_0}(q, \{v_e\}, \{u_x^{(i)}\})$
with $u_x^{(i)} = v_{\<x y_i\>}$.
Since both sides of \reff{eq6.star3} are polynomials in $q$
and the equality holds for infinitely many values of $q$,
it must hold identically.

By applying Theorem \ref{thm6.1} to the graph $G_0$,
we can obtain new results for the
ordinary Potts-model partition function of the graph $G$.
In particular, given {\em any}\/ graph $G=(V,E)$
and any vertex $y \in V$, we can interpret $G$ as the join of
$G_0 \equiv G \setminus y$
(the graph obtained from $G$ by deleting $y$ and all edges incident on it)
and $K_1$.
(Any edge $\< xy \>$ that was not originally present in $G$
 can be introduced and given $v_{\<xy\>} = 0$.)
More generally, given any $N$-clique $y_1,\ldots,y_N$ of $G$,
we can interpret $G$ as the join of
$G_0 \equiv G \setminus \{y_1,\ldots,y_N\}$ and $K_N$;
however, for $N>1$ we must require that $v_{\<y_i y_j\>} = -1$
for each pair $i \neq j$.
Theorem \ref{thm6.1}, Lemma \ref{lemma6.2}(b,c)
and Proposition \ref{prop4.2} then yield
an extension of Corollary \ref{cor5.2}.
To state it, we first define $\widetilde{C} = \widetilde{C}(r,N,\bar{v})$
to be the smallest number for which
\be
   \inf\limits_{\alpha > 0}
   \alpha^{-1} \sum\limits_{n=2}^\infty  e^{\alpha n} \,
       {\widetilde{C}+N \over (\widetilde{C}-N)^n} \,
       \bar{v}^{n-1} \, t_n^{(r)}
   \;\le\;   1
   \;.
\ee
We then have:

\begin{theorem}
\label{thm6.3}
Let $G=(V,E)$ be a loopless finite undirected graph in which all vertices
have degree $\le r$ except perhaps for an $N$-clique $y_1,\ldots,y_N$.
Let $G$ be equipped with complex edge weights $\{ v_e \}_{e \in E}$
satisfying $|1 + v_e| \le 1$ for all $e$
and $v_{\<y_i y_j\>} = -1$ for all $i \neq j$.
Let $v_{max} = \max\limits_{e \in E_0} |v_e|$,
where $E_0$ is the set of edges not incident on any of the vertices
$y_1,\ldots,y_N$.
Then:
\begin{itemize}
   \item[(a)]  All the zeros of $Z_G(q, \{v_e\})$ lie in the disc
      $|q-N| < \widetilde{C}(r,N,v_{max})$.
   \item[(b)]  If, in addition, all the edges $e$ incident on
      any of the vertices $y_1,\ldots,y_N$ satisfy $-1 \le v_e \le 0$,
      then all the zeros of $Z_G(q, \{v_e\})$ lie in the disc
      $|q| < C(r) v_{max} + N$.
\end{itemize}
\end{theorem}

\noindent
And for the chromatic polynomials, we have, using Lemma \ref{lemma6.2}(a):

\begin{corollary}
\label{cor6.4}
Let $G=(V,E)$ be a loopless finite undirected graph
in which all vertices, except perhaps one, have degree $\le r$.
Then all the zeros of $P_G(q)$ lie in the union of the discs
$|q| < C(r)$ and $|q-1| < C(r)$.
In particular, they all lie in the disc $|q| < C(r) + 1$.
\end{corollary}

\noindent
Thus, the zeros of $P_G(q)$ can be bounded in terms of the
{\em second-largest}\/ degree of a vertex in $G$.
Such a result was recently conjectured by Shrock and Tsai \cite{Shrock_99a};
see Section \ref{sec7} for further discussion.

Let us note that the phrase ``except perhaps one'' in Corollary \ref{cor6.4}
{\em cannot}\/ be replaced here by ``except perhaps two'',
not even in the case $r=2$.
Indeed, I have elsewhere \cite{Sokal_hierarchical} constructed
a family of planar graphs in which all but two vertices have degree 2
and whose chromatic roots are together dense in
$\{ q \in \C \colon\;  |q-1| \ge 1 \}$.
Modifications of these graphs show also \cite{Sokal_hierarchical}
that the condition $v_{\<y_i y_j\>} = -1$ for $i \neq j$
in Theorem \ref{thm6.3} (when $N>1$) cannot be relaxed.

Let us now give the proof of Lemma \ref{lemma6.2}.
We will need the following elementary fact:

\begin{lemma}
   \label{lemma6.5}
Let $z$ and $a$ be complex numbers.  Then
\be
   |z + \lambda a|^2  \;\ge\;  |z+a| \, (|z| - |a|)
\ee
whenever $0 \le \lambda \le 1$.
\end{lemma}

\proof
Simple calculus shows that
\be
   \min\limits_{0 \le \lambda \le 1}  |z + \lambda a|^2  \;=\;
   \cases{ |z|^2                 & if $\real(z^* a) \ge 0$  \cr
           \noalign{\vskip 2mm}
           |z|^2 - {\displaystyle \real(z^* a)^2 \over
                    \displaystyle |a|^2}
                                 & if $-|a|^2 \le \real(z^* a) \le 0$ \cr
           \noalign{\vskip 2mm}
           |z+a|^2               & if $\real(z^* a) \le -|a|^2$  \cr
         }
\ee
In the first two cases we clearly have
\be
   \min\limits_{0 \le \lambda \le 1}  |z + \lambda a|^2  \;\ge\;
   |z|^2 - |a|^2  \;=\;  (|z| + |a|) (|z| - |a|)   \;\ge\;
   |z+a| \, (|z| - |a|)
   \;,
\ee
while in the third case we have
\be
   \min\limits_{0 \le \lambda \le 1}  |z + \lambda a|^2  \;=\;
   |z+a|^2   \;\ge\;   |z+a| \, (|z| - |a|)
   \;.
\ee
\qed

\proofof{Lemma \ref{lemma6.2}}
We use the shorthand
$w(S) = \q(S) / \prod\limits_{x \in S} \q(\{x\})$.

(a)  Let $|S| = n$ and suppose that the sequence $(u_x^{(1)})_{x \in S}$
consists of $m$ $-1$'s and $n-m$ 0's.  Then
\be
   w(S)  \;=\;  \cases{ q^{-(n-1)}                  & if $m=0$   \cr
                        \noalign{\vskip 2mm}
                        q^{-(n-m)} (q-1)^{-(m-1)}   & if $1 \le m \le n$ \cr
                      }
\ee
from which \reff{eq_lemma6.2a} immediately follows.

(b)  Let $S = \{x_1,\ldots,x_n\}$;  we then have $\q(\{x_j\}) = q + u_j$
and $\q(S) = q + \bar{u}$ with
$-N \le \bar{u} \le u_1,\ldots,u_n \le 0$.
Now apply Lemma \ref{lemma6.5} with $z=q$, $a = \bar{u}$
and $\lambda = u_j/\bar{u}$ for $j=1,2$:  we have
\be
   \left|  {\q(S) \over \q(\{x_1\}) \, \q(\{x_2\})}  \right|
   \;\le\; 
   {1 \over |q| - |\bar{u}|}  
   \;\le\; 
   {1 \over |q| - N}
\ee
and hence
\be
   |w(S)|  \;\le\;
   {1 \over |q| - N} \prod\limits_{j=3}^n {1 \over |q + u_j|}
   \;,
 \label{eq_lemma6.2b_better}
\ee
which implies \reff{eq_lemma6.2b}.

(c) This bound is trivially obtained by bounding the numerator and
denominator separately.
\qed

\medskip
\par\noindent
{\bf Remarks.}  1.  For simplicity, I have not bothered to exploit
the full strength of \reff{eq_lemma6.2b_better}, which is quite a bit
sharper than \reff{eq_lemma6.2b}.

2.  I am not entirely happy with Lemma \ref{lemma6.2},
and I suspect that it can be improved.
In particular, it is disconcerting that \reff{eq_lemma6.2b}
is not uniformly stronger than \reff{eq_lemma6.2c},
even though the corresponding hypothesis on the $u_x^{(i)}$
{\em is}\/ strictly stronger.


\section{Some Conjectures and Open Questions}   \label{sec7}

The bounds in this paper are, of course, far from sharp,
and it is of some interest to speculate on what the best-possible
results might be.  Let us define
\be
   C_{opt}(r)   \;=\;
   \max\{|q| \colon\;  P_G(q) = 0
       \hbox{ for some loopless graph $G$ of maximum degree $r$} \}
   \;.
\ee
The example of the complete graph $K_{r+1}$ shows that $C_{opt}(r) \ge r$.
It is easy to see that $C_{opt}(1) = 1$ and $C_{opt}(2) = 2$;
and there is some evidence that $C_{opt}(3) = 3$.\footnote{
   Biggs, Damerell and Sands \cite{Biggs_72}
   have verified that the chromatic roots
   of all 3-regular graphs with $\le 10$ vertices,
   as well as those of ladders (``prisms'') and M\"obius ladders
   of arbitrary length, lie in $|q| \le 3$.
   Read and Royle \cite{Read_91} have extended this verification
   to all 3-regular graphs with $\le 16$ vertices,
   as well as to some larger graphs.
}
But, at least for $r \ge 4$,
$C_{opt}(r)$ must in fact be strictly larger than $r$,
as is shown by numerical computations on the
complete bipartite graph $K_{r,r}$
(see Table \ref{table2}).\footnote{
   Recall \cite{Swenson_73,Laskar_75} that
   $$
      P_{K_{m,n}}(q)  \;=\;
         \sum_{k=0}^m S(m,k) \, q_{\<k\>} \, (q-k)^n
      \;,
   $$
   where $S(m,k)$ is the Stirling number of the second kind
   (the number of ways of partitioning a set of $m$ elements
   into $k$ nonempty subsets) \cite[pp.~33--38]{Stanley_86}
   and $q_{\<k\>} = q (q-1) \cdots (q-k+1)$.
   See Woodall \cite[pp.~219--220]{Woodall_77} and Brown \cite{Brown_98a}
   for some properties of the chromatic zeros of the $K_{m,n}$.
}
Indeed, Gordon Royle (private communication) has conjectured
that, for $r \ge 4$, $K_{r,r}$ is the graph of maximum degree $r$
having the largest chromatic roots (in modulus).
It would be useful to have a better understanding of the
chromatic zeros of the complete bipartite graphs $K_{m,n}$.
In particular, it would be useful to have a {\em proof}\/ that
$K_{r,r}$ has chromatic roots of magnitude $>r$ for all $r \ge 4$;
and it would be valuable to understand the asymptotic behavior
of the chromatic roots of $K_{m,n}$ as $m,n \to \infty$
in various ways (e.g.\ with $\alpha = m/n$ fixed).

\begin{table}
\begin{center}
\begin{tabular}{rr@{$\:\pm\:$}r@{$\,i\quad$}r}
\multicolumn{1}{c}{$r$}  &
   \multicolumn{2}{c}{$q$}  & \multicolumn{1}{c}{$|q|$}   \\
\hline \\[-4mm]
  2 &   1.500000 &   0.866025 &   1.732051  \\
  3 &   2.140640 &   1.948682 &   2.894772  \\
  4 &   2.802489 &   3.097444 &   4.177093  \\
  5 &   3.469365 &   4.291184 &   5.518221  \\
  6 &   4.138450 &   5.516667 &   6.896404  \\
  7 &   4.808805 &   6.765768 &   8.300616  \\
  8 &   5.480007 &   8.033190 &   9.724331  \\
  9 &   6.151830 &   9.315289 &  11.163316  \\
 10 &   6.824136 &  10.609446 &  12.614641  \\
 11 &   7.496833 &  11.913711 &  14.076186  \\
 12 &   8.169855 &  13.226591 &  15.546358  \\
 13 &   8.843156 &  14.546915 &  17.023928  \\
 14 &   9.516697 &  15.873744 &  18.507925  \\
 15 &  10.190450 &  17.206318 &  19.997566  \\
 16 &  10.864391 &  18.544006 &  21.492211  \\
 17 &  11.538501 &  19.886280 &  22.991328  \\
 18 &  12.212764 &  21.232697 &  24.494469  \\
 19 &  12.887165 &  22.582876 &  26.001256  \\
 20 &  13.561693 &  23.936489 &  27.511362
\end{tabular}
\end{center}
\caption{
   The chromatic root of largest modulus for the
   complete bipartite graphs $K_{r,r}$ for $2 \le r \le 20$.
}
\label{table2}
\end{table}

Using the Dobrushin uniqueness theorem \cite{Georgii_88,Simon_93},
it can be proven \cite{Salas_97}
that for a countable graph $G$ of maximum degree $r$,
the $q$-state Potts-model Gibbs measure on $G$ is unique
for all integer $q > 2 r$ whenever
$-1 \le v_e \le 0$ for all edges $e$.
Uniqueness of the Gibbs measure is one of several (inequivalent)
notions of ``absence of phase transition'' \cite{Georgii_88,Simon_93}.
It does not imply the analyticity of the free energy,
but it does make it plausible.\footnote{
   Indeed, it was by meditating on possible extensions
   of the theorem in \cite{Salas_97}
   that I was led to conjecture the results in this paper.
}
Likewise, a result that holds for {\em integer}\/ $q > q_0$
need not hold for all {\em real}\/ $q > q_0$, much less for a complex
neighborhood of that real semi-axis; but it does suggest that such a
result might be true.
It is not unreasonable, therefore, to conjecture that there is
a complex domain $D_r$ containing the interval $(2r,\infty)$
of the real axis, such that $Z_G(q, \{v_e\}) \neq 0$
whenever $q \in D_r$, $-1 \le v_e \le 0$ for all edges $e$,
and $G$ has maximum degree $\le r$.
Indeed, it is quite possible that
$D_r = \{ q \colon\; |q| > 2r \}$ works;
this would be a slight extension of the conjecture that $C_{opt}(r) \le 2r$.

We can pose these questions more generally as follows:
Let $\scrg$ be a class of finite graphs,
and let $\scrv$ be a subset of the complex plane.
Then we can ask about the sets
\begin{eqnarray}
   S_1(\scrg, \scrv)   & = &
      \bigcup_{G \in \scrg} \;
      \bigcup_{v \in \scrv} \;
      \{q \in \C \colon\;  Z_G(q,v) \,=\, 0 \}               \\
   S_2(\scrg, \scrv)   & = &
      \bigcup_{G \in \scrg} \;
      \bigcup_{\{v_e\} \colon\; v_e \in \scrv \; \forall e}
      \{q \in \C \colon\;  Z_G(q, \{v_e\}) \,=\, 0 \}
\end{eqnarray}
Among the interesting cases are
the chromatic polynomials $\scrv = \{-1\}$,
the antiferromagnetic Potts models $\scrv = [-1,0]$,
and the complex antiferromagnetic Potts models
$\scrv = A \equiv \{v \in \C \colon\; |1 + v| \le 1 \}$.
Indeed, one moral of this paper is that some questions
concerning chromatic polynomials are most naturally studied
in the more general context of antiferromagnetic or
complex antiferromagnetic Potts models
(with not-necessarily-equal edge weights).
In Corollary \ref{cor5.2} we have shown that
the set $S_2(\scrg_r, A)$ is bounded,
where $\scrg_r$ is the set of all loopless graphs of maximum degree $\le r$;
and in Theorem \ref{thm6.3} we have extended this to $S_2(\scrg'_r, A)$,
where $\scrg'_r$ is the set of all loopless graphs of second-largest
degree $\le r$.
But it would be interesting to examine in more detail the location
of all these sets in the complex plane, and to prove sharper bounds.

Another direction in which the results of this paper could be extended
is by finding a criterion {\em weaker}\/ than bounded maximum degree
(or bounded second-largest degree)
under which the zeros of $P_G(q)$ and $Z_G(q, \{v_e\})$ could be shown
to be bounded.
An interesting idea was suggested very recently by
Shrock and Tsai \cite{Shrock_99a},
who studied a variety of families of graphs
and arrived at a conjecture that can be rephrased as follows:
For $G=(V,E)$ and $x,y \in V$, define
\begin{subeqnarray}
   \lambda(x,y)
      & = &  \hbox{max \# of edge-disjoint paths from $x$ to $y$} \\
      & = &  \hbox{min \# of edges separating $x$ from $y$}
\end{subeqnarray}
and
\be
   \Lambda(G)   \;=\;   \max\limits_{x \neq y}  \lambda(x,y)   \;.
\ee
Clearly $\lambda(x,y) \le \min[\deg(x), \deg(y)]$
and hence $\Lambda(G) \le $ second-largest degree of $G$.
Now let $\scrg^\Lambda_r$ be the set of all loopless graphs
with $\Lambda(G) \le r$.
Then the conjecture is that the set $S_2(\scrg^\Lambda_r, \scrv)$
is bounded, where $\scrv = \{-1\}$ or $[-1,0]$ or perhaps even $A$.\footnote{
   Shrock and Tsai \cite{Shrock_99a} studied only the
   chromatic-polynomial case $\scrv = \{-1\}$,
   and proposed an even stronger result, based on the quantity
   $$
      \Lambda_{\hboxscript{non-adj}}(G)   \;=\;
      \max\limits_{\begin{scarray}
                      x \neq y  \\
                      x,y \hboxscript{ not adjacent}
                   \end{scarray}}
      \lambda(x,y)  \;.
   $$
   But this cannot work for $v \neq -1$:
   a counterexample is obtained \cite{Sokal_hierarchical}
   by gluing together $n$ copies of the cycle $C_k$ (any fixed $k \ge 3$)
   along a single common edge and then taking $n \to \infty$.
}
More generally, one could define
$\lambda(x,y; \{v_e\})$ to be the maximum flow from $x$ to $y$
when $|v_e|$ is taken to be the capacity of edge $e$,
and likewise $\Lambda(G, \{v_e\})$;
this {\em might}\/ lead to the appropriate extension
of Corollary \ref{cor5.5}.
This possible connection of chromatic-polynomial and Potts-model problems
with max-flow problems is intriguing.
Note that $\Lambda(G)$ and $\Lambda(G, \{v_e\})$
possess a ``naturalness'' property that
maximum degree and its relatives lack:
namely, for any graph $G$ with blocks $G_1,\ldots,G_b$,
we have
$\Lambda(G, \{v_e\}) = \max\limits_{1 \le i \le b} \Lambda(G_i, \{v_e\})$;
contrast this with Remark 1 after Corollary \ref{cor5.5}.


\vspace{2cm}
 
\section*{Acknowledgments}
I wish to thank Norman Biggs, Jean Bricmont, Jason Brown, Roberto Fern\'andez,
Ira Gessel, David Jackson, Roman Koteck\'y, Antti Kupiainen,
Neal Madras, Gordon Royle, Jes\'us Salas,
Paul Seymour, Robert Shrock, Gordon Slade, Richard Stanley,
Dave Wagner and Stu Whittington
for valuable conversations and/or correspondence.
I also wish to thank Tony Guttmann, Elliott Lieb and Nick Wormald
for putting me in contact with some of these people.
Finally, I wish to thank an anonymous referee for many helpful comments,
particularly regarding the proofs of
Propositions \ref{prop4.2} and \ref{prop4.5};
and to thank Criel Merino for helping me notice a silly error
in my original proof of Proposition~\ref{prop_penrose}.

This work was begun during a visit to Helsinki
sponsored by the University of Helsinki
and the Finnish National Board of Education.
It was continued during a sabbatical in London
where I had the pleasure of affiliation with
both Imperial College and the London School of Economics.
The final revisions were made during a visiting fellowship at
All Souls College, Oxford.
I wish to thank all these institutions ---
and most especially Antti Kupiainen, Pekka Elo,
Chris Isham, Helena Cronin and John Cardy ---
for their warm hospitality.

This research was supported in part by
U.S.\ National Science Foundation grants PHY--9520978 and PHY--9900769
and by
U.K.\ Engineering and Physical Sciences Research Council grant GR/M 71626.

\end{document}